\documentclass[11pt,a4paper]{article}
\pdfoutput=1
\usepackage{jcappub}
\usepackage[section]{placeins}
\usepackage{bm}
\usepackage{color}
\usepackage{arydshln}

\def\vp{{\varphi}}

\def\cs2{c_{\rm{s}}^2}

\newcommand\calp{\mathcal{P}}

\newcommand\calh{\mathcal{H}}
\newcommand\mpl{m_{\rm Pl}}

\renewcommand\({\left(}
\renewcommand\){\right)}

\newcommand\be{\begin{equation}}
\newcommand\ee{\end{equation}}
\newcommand\bea{\begin{eqnarray}}
\newcommand\eea{\end{eqnarray}}
\newcommand\eq[1]{Eq.~(\ref{#1})}
\newcommand\eqs[2]{Eqs.~(\ref{#1}) and (\ref{#2})}
\newcommand\eqss[3]{Eqs.~(\ref{#1}), (\ref{#2}) and (\ref{#3})}
\newcommand\fig[1]{Fig.~(\ref{#1})}
\newcommand\figs[2]{Figures~(\ref{#1}) and (\ref{#2})}
\newcommand\figss[3]{Figures~(\ref{#1}), (\ref{#2}) and (\ref{#3})}
\newcommand\eps{\epsilon}

\newcommand{\rme}{\mathbf{\rm{e}}}

\begin{document}
\title{Observable Spectra of Induced Gravitational Waves from Inflation}
\author[a]{Laila Alabidi}
\emailAdd{laila@yukawa.kyoto-u.ac.jp}
\author[b,c]{Kazunori Kohri}
\emailAdd{kohri@post.kek.jp}
\author[a]{Misao Sasaki}
\emailAdd{misao@yukawa.kyoto-u.ac.jp}
\author[d]{Yuuiti Sendouda}
\emailAdd{sendouda@cc.hirosaki-u.ac.jp}

\affiliation[a]{Yukawa Institute for Theoretical Physics, Kyoto University, Kyoto 606-8502, Japan}
\affiliation[b]{Cosmophysics Group, Theory Center, IPNS, KEK, Tsukuba 305-0801, Japan}
\affiliation[c]{The Graduate University for Advanced Study (Sokendai), Tsukuba 305-0801, Japan}
\affiliation[d]{Graduate School of Science and Technology, Hirosaki University, Hirosaki, Aomori 036-8561, Japan}


\abstract{
Measuring the primordial power spectrum on small scales is a powerful tool in inflation model building, yet constraints
from Cosmic Microwave Background measurements alone are insufficient to place bounds stringent enough to be
appreciably effective. For the very small scale spectrum, those which subtend angles of less than $0.3$ degrees on the sky,
an upper bound can be extracted from the astrophysical constraints on the possible production of primordial black holes in the early universe.
A recently discovered observational by-product of an enhanced
power spectrum on small scales, induced gravitational waves, have been shown to be within the range of proposed
space based gravitational wave detectors; such as NASA's LISA and BBO detectors, and the Japanese DECIGO detector.
In this paper we explore the impact such a detection would have on models of inflation known to
lead to an enhanced power spectrum on small scales, namely the Hilltop-type and running mass models. 
We find that the Hilltop-type model can produce observable induced gravitational waves within
the range of BBO and DECIGO for integral
and fractional powers of the potential within a reasonable number of $e-$folds. We also find that the running 
mass model can produce a spectrum within the range of these detectors, but require that inflation terminates after 
an unreasonably small number of $e-$folds. Finally, we argue that if the thermal history of the Universe
were to accomodate such a small number of $e-$folds the Running Mass Model can produce Primordial Black Holes
within a mass range compatible with Dark Matter, i.e. within a mass range $10^{20} {\rm g}\lesssim M_{\rm BH}\lesssim10^{27}{\rm g}$.}
\date{\today}
\maketitle

\section{Introduction}
One of the main goals of cosmology is to uncover a unique model of
inflation, an important achievement that will lead to a fuller
understanding of the dynamics of the very early universe and allow us
to better compare with fundamental theory.  To fully probe the
inflationary potential, an accurate measure of the primordial spectrum
on all scales is required.  To date, Cosmic Microwave Background (CMB)
experiments such as COBE and WMAP \cite{Komatsu:2010fb}, Baryon
Acoustic Oscillations (BAO) in Galaxy Survey experiments (such as SDSS
\cite{Abazajian:2008wr}) and  Super Novae (SN) \cite{Hicken:2009dk}
observations have constrained the spectrum on large scales to
$\mathcal{P}_\zeta=2.325\times10^{-9}$ with an uncertainty of
$\pm0.098\times10^{-9}$.  Here, by large scale we mean scales which
subtend angles on the sky of more than $\theta=0.3^{\circ}$.  Bounds
on the smaller scale spectrum arise from such sources as the
Lyman-$\alpha$ forest which measures this spectrum via  a trace of the
Baryonic power spectrum from the intergalactic medium
(e.g. Ref.~\cite{Meiksin:2007rz})  \footnote{See
Ref.~\cite{Bird:2010mp} for the latest analysis of the Lyman-$\alpha$
forest}, weak lensing (e.g. Ref.~\cite{Lewis:2006fu}), the
Sunyaev-Zeldovich effect (e.g. Ref.~\cite{Komatsu:2002wc}),  bounds from
Ultra compact Mini Halos
\cite{Josan:2010vn,Bringmann:2011ut,Yang:2011eg,Li:2012qh,Scott:2012kx} and the astrophysical bounds
on the production of Primordial Black Holes (PBHs)
\cite{Carr:2009jm,Josan:2009qn}. 

The PBH bound constrains the spectrum on the smallest scales, those which exit the horizon towards the end of inflation, and 
it is the uncertainty in this bound which allows for some leeway in inflation model building. 
In previous work \cite{Leach:2000ea, Kohri:2007gq, Kohri:2007qn, Alabidi:2009bk}
it has been shown that Hilltop-type models and their ilk can give rise to an
enhanced spectrum towards the end of inflation, within the observational bounds of PBH production.
An actual measurement of this feature would prove decisive in inflation model discrimination 
\footnote{For example, Ref.~ \cite{Josan:2010cj} uses the PBH bound to constrain the cosmological observables
}. References \cite{Matarrese:1993zf,Matarrese:1997ay,Noh:2004bc,Carbone:2004iv,
Nakamura:2004rm,Ananda:2006af, Baumann:2007zm}
have shown that the primordial scalar spectrum can result in the 
production of what are known as induced gravitational waves \footnote{See also
Ref.~\cite{Cook:2011hg} for an alternative scenario, where gravitational waves are induced
by particle production during inflation}. These gravitational waves have an enhanced energy density for 
an enhanced scalar spectrum which can exceed that of the primordial tensor spectrum for small field models of inflation. 
This opens up another avenue for fully probing the primordial spectrum and placing constraints on the 
cosmological and inflationary parameters (e.g. Ref~\cite{Assadullahi:2009jc}).

In  this scenario, scalar perturbations re-enter the horizon during Big Bang Nucleosynthesis perturbing the background
metric and inducing a tensor perturbation. The modes which enter the horizon during
radiation domination generate gravitational waves on scales accessible
to space based gravitational wave detectors \cite{Baumann:2007zm,Assadullahi:2009jc} 
such as LISA \cite{lisa}, BBO and DECIGO \cite{Seto:2001qf,decigo}.
The scalar perturbations which enter the horizon during matter domination
generate an induced gravitational wave spectrum accessible to CMB experiments \cite{Mollerach:2003nq}, and
since these waves do not redshift, their spectrum is scale dependent \cite{Baumann:2007zm}, 
making them distinguishable from their primordial, scale independent, brethren. 
We are interested in the former scales, which leave the horizon towards the end of inflation,
and re-enter during the radiation era. 

Previous work \cite{Saito:2009jt, Bugaev:2009kq,Bugaev:2010bb} calculated
the induced gravitational wave spectra
for step and power law primordial spectra, and 
reference \cite{Bugaev:2009zh} calculated the induced GW-spectrum of the running mass model. 
In this work we calculate the spectra of induced gravitational waves for various parameter 
choices in both the running mass and Hilltop-type models. 
We find that in fact, the Hilltop-type model can result in a significant amplitude of gravitational waves, measurable
by the BBO/DECIGO and cross-correlated DECIGO detectors. 

In Section \ref{sec:IGW} we briefly present the equations for Induced Gravitational Waves, then in Section \ref{sec:PBH} we evaluate the upper bounds on the scalar spectrum 
from the possible production of Primordial Black Holes. In Section \ref{sec:inf} we summarise the inflationary parameters and their observational bounds. In Section \ref{models} we 
present the models of inflation that we are analysing and their parameters. In Section \ref{sec:results}
we present our results with some discussion. Finally, in Section \ref{sec:conclusion} we summarise the main results of this work.

The following conventions are utilised in this paper: $\tau$ refers to conformal
time and is related to proper time $t$ as $d\tau=dt/a$, $a$ is the scale factor, 
and the conformal
Hubble parameter $\calh$ is related to the Hubble parameter $H\equiv\dot{a}/a$ as $\calh=aH$.
Scales are denoted by $k$, are given in units of inverse megaparsec $\rm{Mpc}^{-1}$
and are related to physical frequency $f$ as $f=ck/(2a\pi)$ where $c$ is the speed 
of light. We assume a radiation dominated universe at the time of the formation 
of the gravitational waves, in which case we have $a=a_0(\tau/\tau_0)$, $\calh=\tau^{-1}$,
and the scale at re-entry is $k=\tau^{-1}$.

\section{Induced Gravitational Waves}
\label{sec:IGW}

After Inflation has ended, scalar perturbations begin the re-enter the horizon and interact in such a a way as to induce gravitational waves. These gravitational waves have a spectrum that is dependent on scale and is given as 
\be
P_h(k)=\frac{1}{a^2}\int_0^\infty{}d\tilde{k}\int_{-1}^{1}d\mu\frac{k^3\tilde{k}^3}{|\mathbf{k}-\tilde{\mathbf{k}}|^3}(1-\mu^2)^2P(|\mathbf{k}-\tilde{\mathbf{k}}|)P(\tilde{k})I_1(k,\tilde{k},\tau)I_2(k,\tilde{k},\tau)\,
\ee
derived in appendix (\ref{app:PGW}). Where $P(k)$ is the scalar spectrum given by the model of inflation, 
$\mu$ is the cosine of the angle between the two modes and $I_1$ and $I_2$ are the
time integrals given in appendix (\ref{app:PGW}). In this paper we work with the parameters $v=\tilde{k}/k$,
$y=\sqrt{1+v^2-2v\mu}$ and $x=k\tau$, in which case the spectrum of induced gravitational waves takes the form
\be
P_h(k)=\frac{k^2}{x^2}\int_0^\infty{}dv\int_{|v-1|}^{|v+1|}dy\frac{v^2}{y^2}(1-\mu^2)^2P(kv)P(ky)\tilde{I}_1\tilde{I}_2 .
\label{eq:Ph}
\ee
What is important to note here is that the time integrals, given in \eq{tauint} 
are independent of the model of inflation, and depend only on the epcoh of evaluation.
In this paper we work only with gravitational waves induced during the radiation era, and therefore the time integrals are given 
by \eqs{t1}{t2}. Their behaviour is plotted in \fig{tau_int},
and as is clear, the envelope of the integrals is constant for a fixed scale and rapidly decay to zero for $\tilde{k}\gg k$.

Since gravitational wave detectors will measure the amplitude of the energy density of gravitational waves, we will
be presenting our results in terms of the dimensionless variable $\Omega_{GW}$ which defines the variation of the energy
density with respect to the logarithm of the scale. We discuss this parameter more in appendix (\ref{app:omega}), for
now we present the form of $\Omega_{GW}$ for scales which re-enter the horizon during the radiation era

\be
\Omega_{GW}=\frac{1}{1+z_{eq}}P_{h}(k)
\label{eq:omega}
\ee
where $eq$ denotes matter-radiation equality.
\section{Primordial Black Holes}
\label{sec:PBH}

In this section we evaluate upper bounds on $\calp_\zeta$ from  PBH
formation, which is severely constrained by various astrophysical and
cosmological observations~\cite{Carr:2009jm}.

In the simplest model, PBHs are formed due to instantaneous gravitational collapse during the 
radiation-dominated era. Their mass is given in terms of the energy within the sound horizon at the time 
of formation, $ M \equiv c^3/(2\,G\,H(t)) $\,, as $ M_\mathrm{BH} = \gamma\,M $\,, where $ \gamma $ is a numerical factor to represent uncertainty.
The corresponding comoving wave number of a density perturbation is
\begin{equation}
k_\mathrm{BH}
= a\,H
\simeq
  \frac{k_\mathrm{eq}}{2^{1/4}}\,
  \left(\frac{g_*}{g_{*\mathrm{eq}}}\right)^{-1/12}\,
  \left(\frac{M}{M_\mathrm{eq}}\right)^{-1/2}
\quad
(M \ll M_\mathrm{eq})\,,
\label{eq:k}
\end{equation}
where
\begin{equation}
k_\mathrm{eq}
\equiv
  a_\mathrm{eq}\,H_\mathrm{eq}
\approx
  0.00974\,\mathrm{Mpc}^{-1}\,,
\quad
M_\mathrm{eq}
\equiv
  \frac{c^3}{2\,G\,H_\mathrm{eq}}
\approx
  6.67 \times 10^{50}\,\mathrm g\,.
\end{equation}

For a Gaussian-distributed density perturbation \footnote{Refer to Refs.~\cite{Bullock:1996at,Ivanov:1997ia,PinaAvelino:2005rm,Chongchitnan:2006wx,Hidalgo:2007vk,Bugaev:2011wy}
for the alternative
scenario where a non-gaussian
distributed density perturbation is considered.}, the energy fraction going into collapsed objects, 
$ \beta(M) $\,, is related to the mass variance at horizon entry, $ \sigma(M)^2 $ via the formula
\begin{equation}
\beta(M)
= \gamma\,\frac{2}{\sqrt{2\,\pi}\,\sigma(M)}
  \int_{1/3}^1 \!\mathrm d\delta\,
  \exp\left(-\frac{\delta^2}{2\,\sigma(M)^2}\right)
\approx
  \gamma\,\mathrm{erfc}\left(\frac{1/3}{\sqrt 2\,\sigma(M)}\right)\,,
\label{eq:beta}
\end{equation}
where the prefactor $ 2 $ embodies the Press--Schechter prescription.
Equation~\eqref{eq:beta} is numerically inverted to give the value of $ \sigma $ for a given upper limit of $ \beta $\,.

\begin{figure}[tb]
\begin{center}
\includegraphics[scale=0.4]{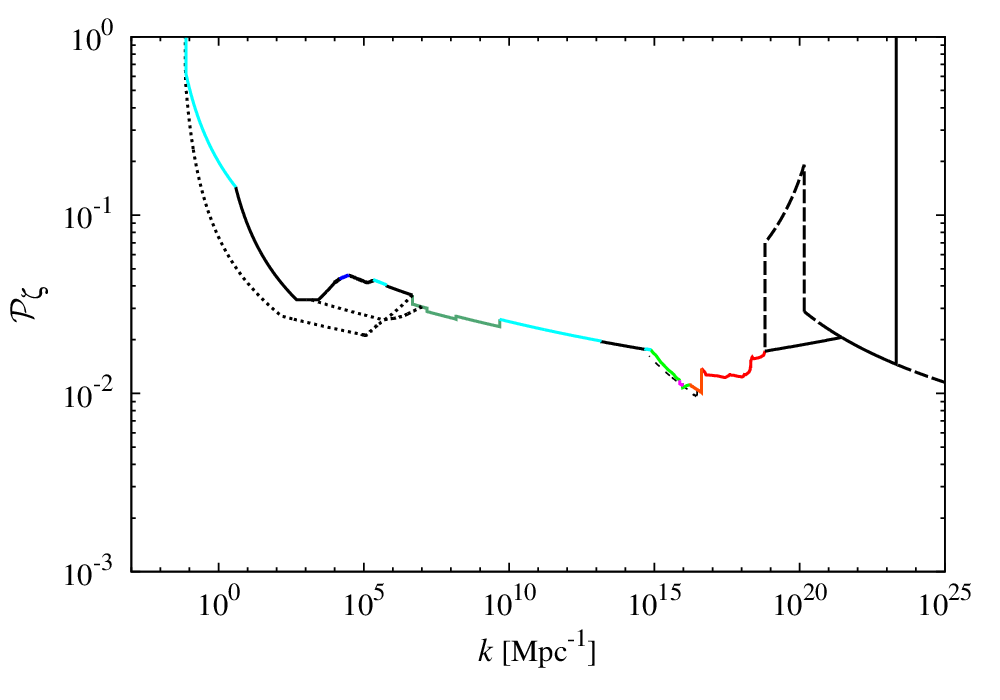}\end{center}
\caption{\label{fig:limit}
Upper limits on $ \mathcal P_\zeta $ from PBH constraints.
For simplicity, $ \gamma $ is set to be unity.
The different coloured solid lines correspond to bounds from different phenomena and experiments as summarised in table~\ref{tab:pbh}.
For comparison, the dotted lines are included to indicate CMB constraints on accreting PBHs;
the thin-dashed line illustrates a potential constraint from future $ 21\,\mathrm{cm} $ line experiments;
the thick-dashed line is the limit necessary to avoid excessive generation of entropy and Planck mass relics 
($ k \lessgtr 2.1 \times 10^{23}\,\mathrm{Mpc}^{-1} $). For the latter the reheating temperature is assumed to be higher than $ 10^{16}\,\mathrm{GeV} $.
See \cite{Carr:2009jm} for details.
}
\end{figure}

The mass variance evaluated at horizon entry (when $ a\,H = k_\mathrm{BH} $ holds) is
\begin{equation}
\sigma(M)^2
= \frac{16}{81} \int_0^\infty\!
  W(k/k_\mathrm{BH})^2\,
  T(k)^2\,
  \left(\frac{k}{k_\mathrm{BH}}\right)^4\,
  \mathcal P(k)\,
  \frac{\mathrm dk}{k}\,,
\end{equation}
where $ W $ is the window function, $ T $ is the transfer function and $ \mathcal P $ is the power spectrum of the curvature perturbation in the comoving gauge.
We choose a Gaussian window function $ W(x) = \mathrm e^{-x^2/2} $ and assume $ T(k) = 1 $ for simplicity.
Since the window function has a sharp cutoff, we only need 
to evaluate the power spectrum around 
$ k_\mathrm{BH} $ to calculate the mass variance on the relevant scale.
Incidentally, the comoving curvature perturbation is 
identical to 
the curvature perturbation in the uniform-density gauge 
$ \zeta $ (up to a sign difference) on superhorizon scales
. Thus $ \mathcal P $ in the above equation can 
simply be replaced by the spectrum of $ \zeta $\,, 
denoted by $ \mathcal P_\zeta $\,.
If the spectrum is locally scale invariant, i.e. $ n(k) \approx 1 $ in the neighbourhood of $ k_\mathrm{BH} $\,, 
as is generally expected in various inflationary models, then the mass variance is estimated as
\begin{equation}
\sigma(M)^2
\approx
  \frac{8}{81}\,\mathcal P_\zeta(k_\mathrm{BH})\,.
\end{equation}
Using this equation and Eq.~\eqref{eq:beta}, we can estimate the upper
limit of $ \mathcal P_\zeta $ from that on the fraction $ \beta(M) =
\beta(M_\mathrm{BH}/\gamma) $; which is given in Fig.~9 of
Ref.~\cite{Carr:2009jm}. In Fig.~(\ref{fig:limit}), we plot the obtained
upper limit on $ \mathcal P_\zeta $ as a function of $k$. 

\begin{table}[tb]
\caption{\label{tab:pbh}
Upper limits on PBH formation from various observations.
Most of them come from non-detections (NDs) of astrophysical phenomena or effects which would be detectable if there were a sufficient number of PBHs.
See \cite{Carr:2009jm} for details. Note that the structure on scales less than 
$10^4[Mpc^{-1}]$ may be further constrained
in the future by the upcoming PIXIE experiment \cite{Chluba}.}
\begin{center}
\begin{tabular}{c|l}
\hline\hline
\shortstack[l]{Wavenumber\\$ \gamma^{1/2}\,k\,[\mathrm{Mpc}^{-1}] $}
& Constraints \\
\hline
$ < 3.9 $
& Density of PBHs ($ \Omega_\mathrm{PBH} $) $ < 0.25 $ \\
$ 3.9\,\times 10^0\text{--}1.5\,\times 10^4 $
& No excessive dynamical friction in the Galactic halo \\
$ 1.5\,\times 10^4\text{--}3.0\,\times 10^4 $
& ND of Poisson fluctuations in Lyman-$ \alpha $ forest \\
$ 3.0\,\times 10^4\text{--}2.1\,\times 10^5 $
& ND of wide binary disruption in the Galaxy \\
$ 2.1 \times 10^5\text{--}6.1 \times 10^5 $
& $ \Omega_\mathrm{PBH} < 0.25 $ \\
$ 6.1 \times 10^5\text{--}4.7 \times 10^6 $
& ND of microlensed quasars \\
$ 4.7 \times 10^6\text{--}4.7 \times 10^9 $
& Lack of MACHO events \\
$ 4.7 \times 10^9\text{--}1.5 \times 10^{13} $
& $ \Omega_\mathrm{PBH} < 0.25 $ \\
$ 1.5 \times 10^{13}\text{--}4.7 \times 10^{14} $
& ND of femto/picolensed gamma-ray bursts \\
$ 4.7 \times 10^{14}\text{--}7.5 \times 10^{14} $
& $ \Omega_\mathrm{PBH} < 0.25 $ \\
$ 7.5 \times 10^{14}\text{--}7.5 \times 10^{15} $
& ND of extragalactic gamma-rays (EGR) \\
$ 7.5 \times 10^{15}\text{--}9.2 \times 10^{15} $
& ND of Galactic gamma-rays \\
$ 9.2 \times 10^{15}\text{--}1.6 \times 10^{16} $
& ND of EGR \\
$ 1.6 \times 10^{16}\text{--}4.2 \times 10^{16} $
& No damping of small-scale CMB anisotropy \\
$ 4.2 \times 10^{16}\text{--}6.4 \times 10^{18} $
& Standard BBN \\
$ 6.4 \times 10^{18}\text{--}2.8 \times 10^{21} $
& Density of the lightest SUSY particle (LSP) $ < 0.25 $ \\
& (assuming $ 100\,\mathrm{GeV} $ for LSP's mass) \\
$ 2.8 \times 10^{21}\text{--}2.1 \times 10^{23} $
& Density of Planck mass relics $ < 0.25 $ \\
& (assuming $ \mathcal O(10^{16}\,\mathrm{GeV}) $ for the reheating temperature) \\
\hline\hline
\end{tabular}
\end{center}
\end{table}

\section{Inflationary Parameters}\label{sec:inf}

We consider only single-field canonical models of inflation, where the accelerated expansion 
of the universe is driven by a flat potential, and a slowly changing
scalar field (slowly rolling inflaton).
To parametrize models of inflation, slow roll parameters \cite{Lyth:2009zz} are used,
defined as:

\bea\label{SR}
\eps&=&\frac{\mpl^2}{2}\(\frac{V_{,\vp}}{V}\)^2\nonumber\\
\eta&=&\mpl^2\frac{V_{,\vp\vp}}{V}\nonumber\\
\xi^2&=&\mpl^4\frac{V_{,\vp}V_{,\vp\vp\vp}}{V^2}
\eea
where $V$ is the potential, and
derivatives are with respect to the inflaton field $\vp$. From this we can write down the observational parameters,
the spectral index $n_s$, the running of the spectral index $n_s'$ and
the scalar spectrum $\mathcal{P}_\zeta$:

\bea\label{para}
n_s&=&1+2\eta-6\eps\nonumber\\
n_s'&=&16\eps\eta-24\eps^2-2\xi^2\nonumber\\
\mathcal{P}_\zeta&=&\frac{1}{24\pi^2\mpl^4}\frac{V}{\eps}
\eea
where we neglect to mention the signature of primordial gravitational waves, the tensor to scalar
ratio, since we are only considering small field models and as such this parameter is
negligibly small. 

The next ingredient required is the number of $e-$folds, the logarithmic
ratio of the scale factor at two different times, in this case between the end
of inflation and the time of horizon exit. This is related
to the potential in the slow roll limit as:

\be\label{NSR}
N\simeq\mpl^{-2}\int_{\varphi_e}^{\varphi_*}\frac{V}{V'}d\varphi
\ee
and to the corresponding horizon exiting scale as \cite{Lyth:2009zz}:
\be\label{Nk}
N(k_0)-N(k)=\ln\(\frac{0.002}{k}\).
\ee
where $k_0=0.002\rm{Mpc}^{-1}$ is the pivot scale, and in this paper we effectively take $N(k_0)=0$.

We use the latest data release from the WMAP mission \cite{Komatsu:2010fb}, for the
WMAP data combined with BAO and SN data with a null tensor prior. This
gives us the following bounds on the spectral index, and running of the spectral
index at the $2\sigma$
confidence limit:

\bea
-0.093<1-n_s<0.076\nonumber\\
-0.061<n_s'<0.017
\eea
in this paper we take $n_s=0.95$ or $n_s=0.96$.

\section{The Models of Inflation}\label{models}
\begin{figure}
 \centering\includegraphics[width=\linewidth,totalheight=2in]{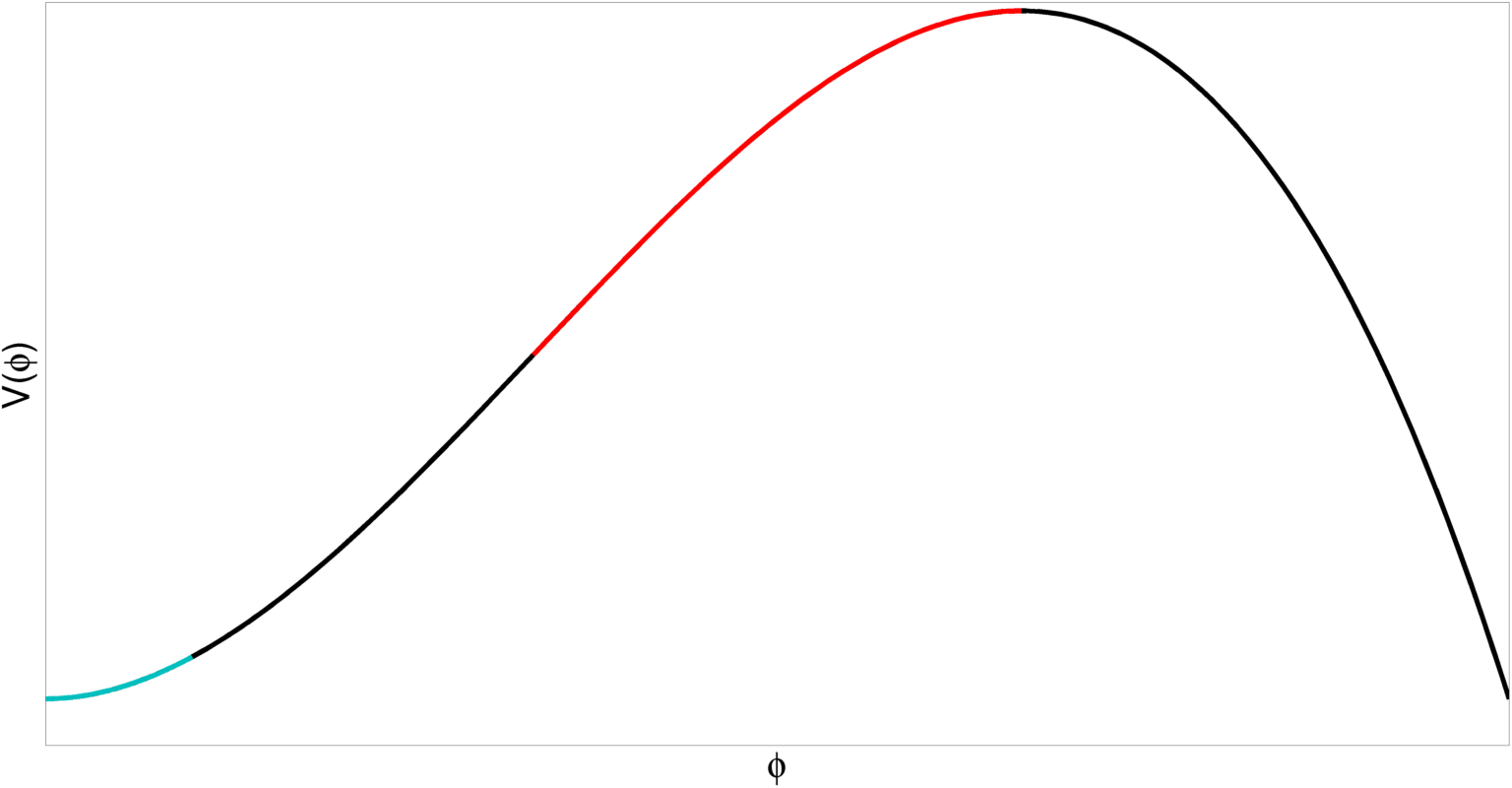}
\caption{An illustration of the Hilltop-type and running models. In our scenario scales of cosmological
interest during the hilltop regime (indicated with red), and the end of inflation occurs once the inflaton
has reached a flatter region of the potential (indicated with blue).}
\label{fig:potential}
\end{figure}

In this paper we analyse two models of inflation, the Hilltop-type model
and the running mass model. Both models have a hilltop regime, an inflection 
point to one side of the hilltop and a steep slope to the other side. In the set-up
of interest, scales of cosmological interest leave the horizon while the inflaton
is on the inflection point side. The inflaton then proceeds to roll down the potential,
past the inflection point and towards a region of further flatness, $\eps\to0$, as illustrated
in fig.~\ref{fig:potential}.
This means that our spectrum
at the pivot scale will satisfy the WMAP bounds while still increasing on
the smaller scales. Ref.\cite{Drees:2011yz} analyse various models of inflation and
conclude that only the running mass model can allow for PBH formation, and we 
note that the hilltop
model is phenomenologically equivalent to the running mass model.
We begin by introducing the Hilltop-type model and the associated scalar spectrum
and then move onto the running mass model and its scalar spectrum. We do
not explicitly define a mechanism for the end of inflation or for the necessary
subsequent reheating. We demand that inflation
is ended abruptly after a specified number of $e-$folds, e.g.
terminated by a waterfall field,  and that reheating is instant.

\subsection{Hilltop-type model}\label{sec:hilltop}

The phenomenological form of the potential \cite{Kohri:2007gq} is
given as:
\be\label{hilltop-potential}
V=V_0\(1+\eta_p\varphi^p-\eta_q\varphi^q\)\,
\ee
where $\eta_p$ and $\eta_q$ are referred to as the mass-coupling terms
and self coupling powers $p$ and $q$ are required to satisfy
$p<q$. This condition is to guarantee the hilltop form of the
potential. Such a potential form also appears in supergravity
models
(e.g. ref.~\cite{Allahverdi:2006cx,Kohri:2007gq,Lin:2008ys,Lin:2009yt,Kohri:2010sj,Hotchkiss:2011gz}).  The model has four
degrees of freedom, with only weak constraints from fundamental
theory. Therefore for each $\{p,q\}$ combination we scan the
$\{\eta_p,\eta_q\}$ parameter space as follows

\begin{itemize}
\item We pre-set both $\eta_p$ and $\eta_q$ to be less than one.
 \item From the range of values $\{0,1\}$, the parameter range of $\{\eta_p,\eta_q\}$ is reduced by requiring that, 
at the pivot scale, the spectral index and the running of the spectral index are within the WMAP bounds, $n_s=0.95$ and $n_s'<0.017$.
\item The parameter range is further reduced by rejecting mass-coupling combinations 
for which the field value at horizon exit is greater than the Planck scale; i.e. we demand $\phi_*<\mpl$.
\item The range is then reduced to unique values of $\{\eta_p,\eta_q\}$ for each model by
demanding that, after N $e-$folds of inflation, the spectrum is close to
but still less than the PBH bound at that $e-$fold,
$\calp(N)\lesssim\calp_{\rm PBH}$. The exception to this is the $N=65$ case, for which we refer
the reader to the following paragraph.
\end{itemize}
The fourth step essentially selects the unique mass-coupling values which maximize the spectrum at the end of inflation, 
compatible with the PBH bound. As we mentioned, the exception to the requirement in the fourth step
is the case where inflation terminates at $N=65$. Since the PBH bound only applies up $N\sim60(k\sim10^{23})$,
we maximise  the spectrum at the intermediate scale and then allow the model to continue to evolve until $N=65$.
Clearly the spectra at $N=60$ for the $N=60$ and $N=65$ model will be the same, the difference is that for
the $N=60$ model $\calp_\zeta(N>60)=0$ while for the $N=65$ model $\calp_\zeta(N>60)\neq0$. A-priori
we expect the final results for the GW spectrum to be the same in the range of interest, since the time integra asymptote to 
zero for $k>\tilde{k}$, but we evaluate the $N=65$ case anyway as a consistency check of our numerics.
We tabulate the parameters and predictions for $n_s'$ and $V_0$ in Appendix~(\ref{app:hill_coeff}) and plot the results for
the first order spectrum for $N=55$, $N=60$ and $N=65$ in \figss{spectra_hill1}{spectra_hill2}{spectra_hill3}. 
\begin{figure}
\centering\includegraphics[width=4in,totalheight=4in]{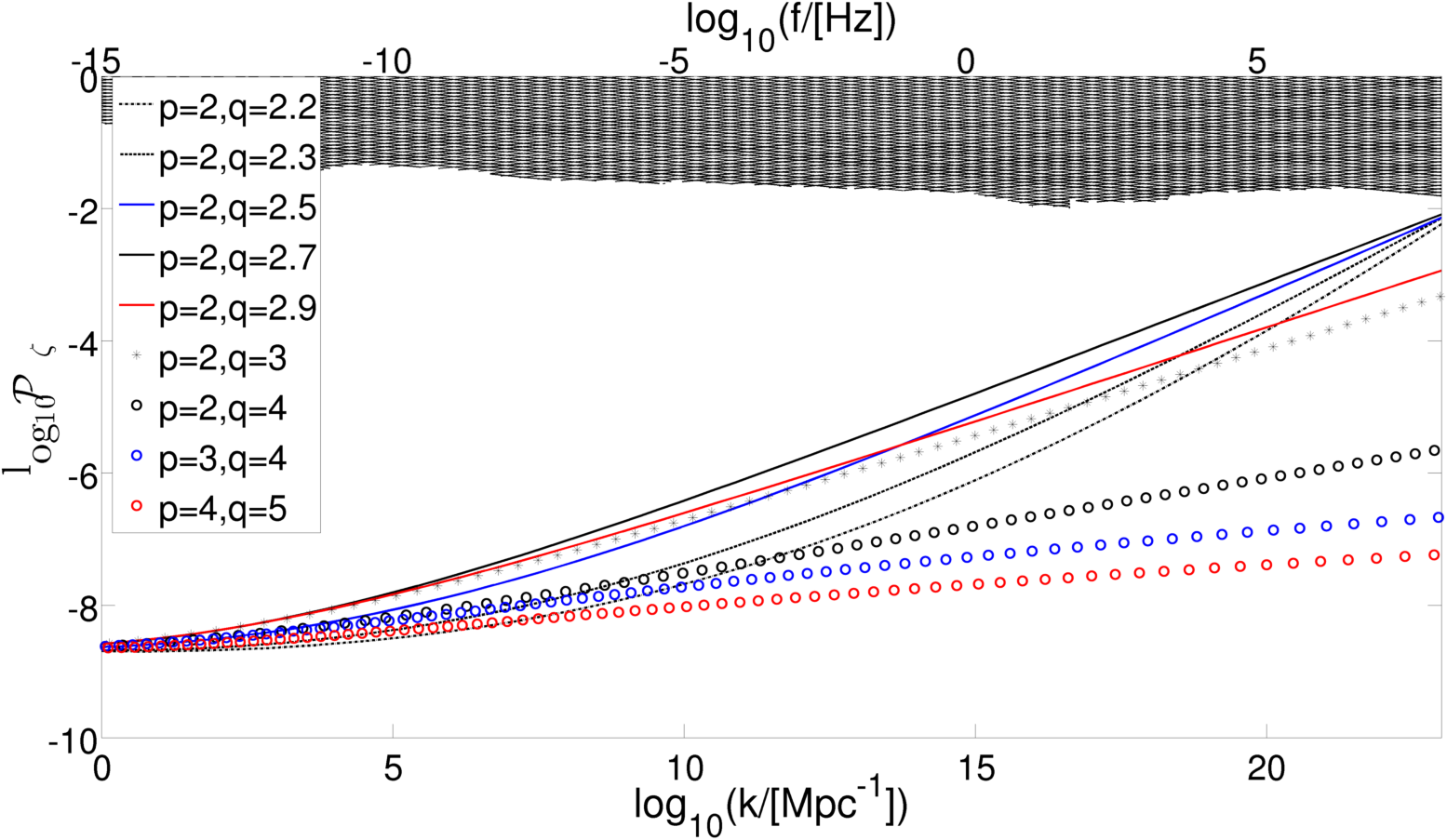}
\caption{The scalar spectra of the hilltop model terminating at $N=65$ ($\log_{10}(k/(0.002[\rm{Mpc}^{-1}]))\sim28.2$) while ensuring that $\calp(N=60)$ is less than the PBH bound
at that scale.  The cross-hatched region is the PBH constraint.}
\label{spectra_hill3}
\end{figure}
\begin{figure}
 \centering\includegraphics[width=4in,totalheight=4in]{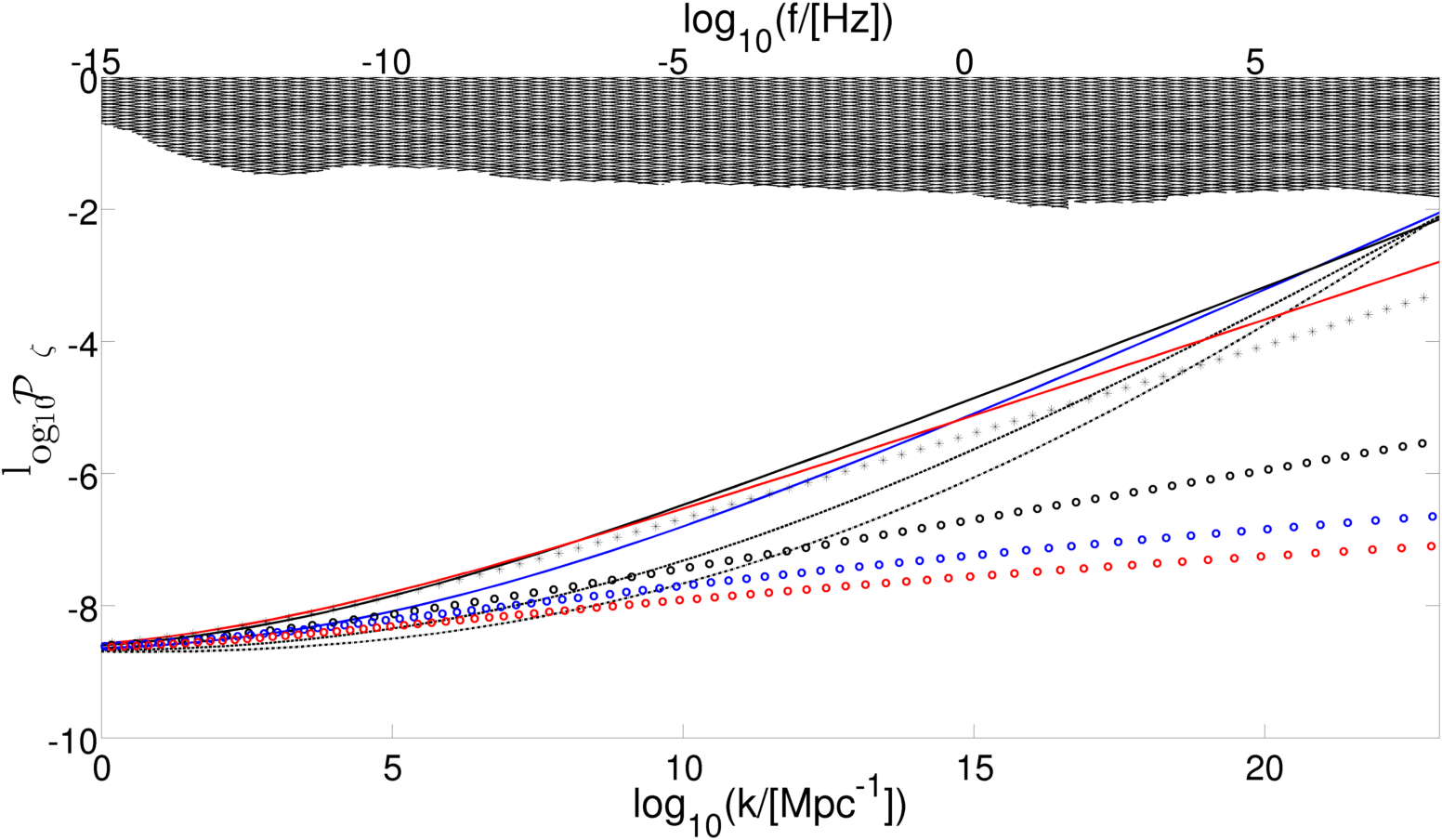}
\caption{Scalar spectra from hilltop inflation with self-coupling powers labelled in the legend 
in \fig{spectra_hill3}. The plot corresponds to maximising the spectrum at $N=60$ ($\log_{10}(k/(0.002[\rm{Mpc}^{-1}]))\sim26)$.
 The cross-hatched region is the PBH constraint.}
\label{spectra_hill2}
\end{figure}
\begin{figure}
\centering
\includegraphics[width=4in,totalheight=4in]{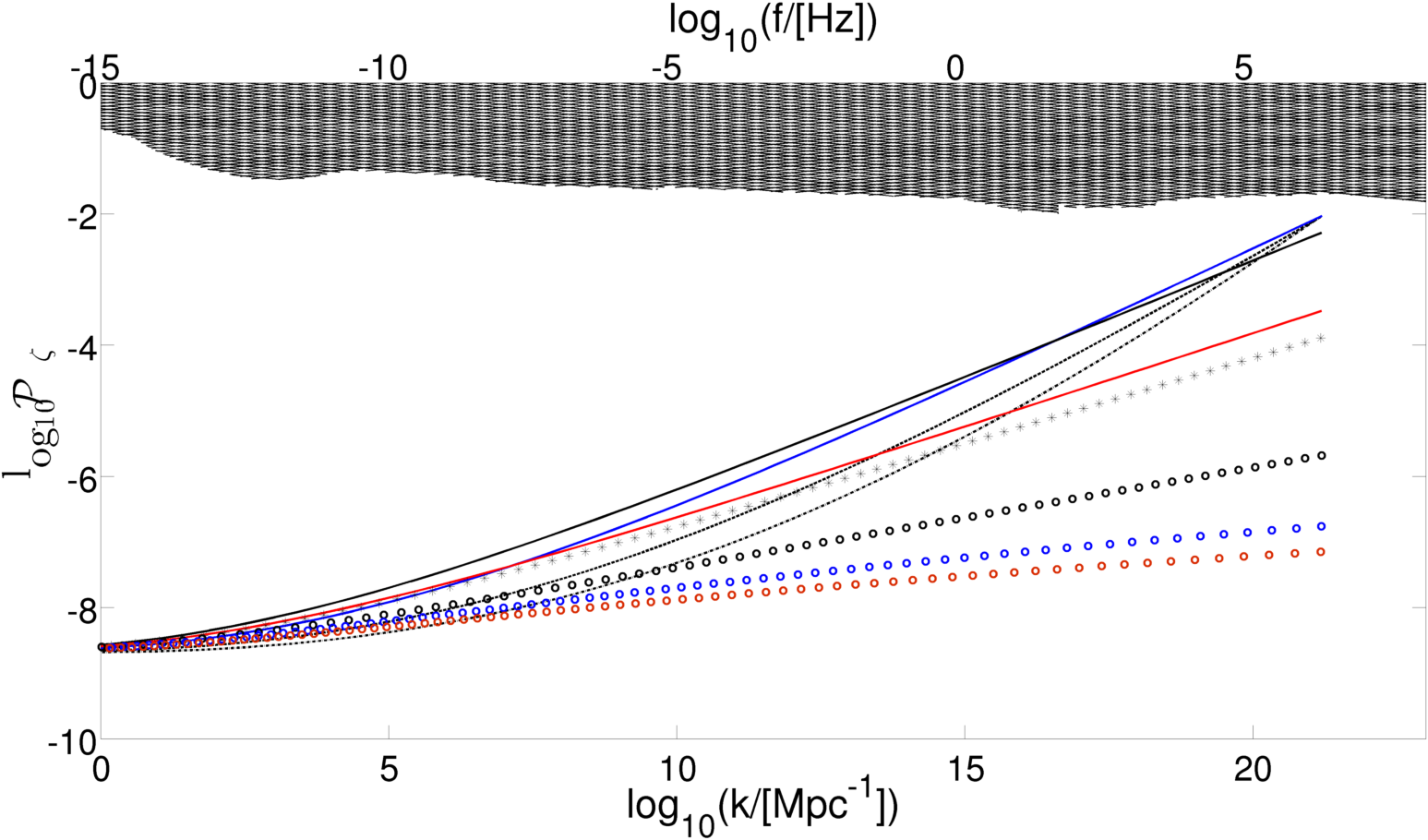}
\caption{Scalar spectra from hilltop inflation with self-coupling powers labelled in the legend 
in \fig{spectra_hill3}. The plot corresponds to maximising the first order
spectrum at $N=55$ ($\log_{10}(k/(0.002[\rm{Mpc}^{-1}]))\sim23.9$), and the cross hatched region is the PBH bound}
\label{spectra_hill1}
\end{figure}

\subsection{The Running Mass Model}

This model \cite{Stewart:1996ey,Covi:1998jp, Covi:1998mb, Lyth:2000qp,Leach:2000ea,Covi:2000qx, Covi:2002th,Covi:2004tp}
is a $\vp^2$ model only with
a mass term which varies with $\vp$. The induced gravitational wave spectrum in this model
was originally evaluated in Ref.~\cite{Bugaev:2009zh}. The potential is of the form:
\be
\frac{V}{V_0}=1-\frac{B_0}{2}\vp^2+\frac{A\vp^2}{2(1+\alpha\ln(\vp))^2}~.
\ee

\begin{figure}
 \centering\includegraphics[width=4in,totalheight=4in]{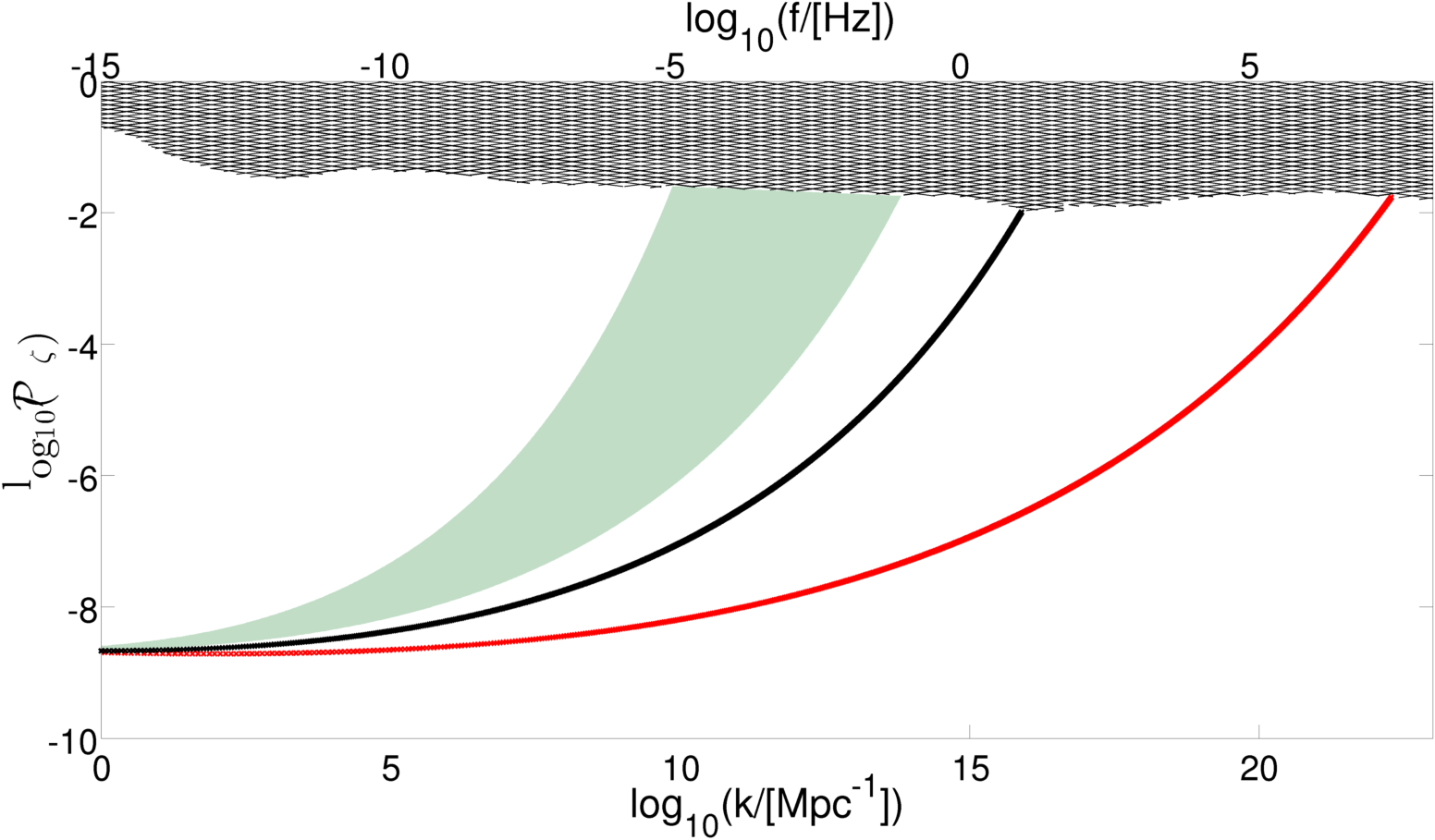}
\caption{ Spectra for the running mass models all satisfying $n_s=0.96$. The 
$n_s'=0.005$ model (black crosses) requires a termination of inflation at
$N\sim43$ ($\log_{10}(k/(0.002[\rm{Mpc}^{-1}]))\sim16$) 
for compatibility with the PBH bound,
and the $n_s'=0.002$ model (red crosses) requires only a termination at $N\sim64$. The hatched line is the PBH bound 
and both parameter combinations lead to the production of PBHs towards the end of inflation. The green shaded
region corresponds to range of parameters in the running mass model which
result in the production
of PBHs whose energy density agrees with that of Dark Matter. The range corresponds to $10<\log_{10}(k/[\rm{Mpc}^{-1}])<14$,
where the lower $k$ value represents $n_s'=0.012$,
and inflation terminating at $N=29$ while
the upper $k$ value represents $n_s'=0.0067$, 
and inflation
terminating at $N=38.5$.}
\label{spectra_rmm}
\end{figure}

We know from previous work \cite{Alabidi:2009bk} that the parameter values $A=2.4$, $B_0=2.42$, $\alpha=0.01$ 
satisfy WMAP bounds and lead to PBHs at the end of
inflation. 
In this case, for $n_s=0.95$,
then  
$V_0^{1/4}=0.003\mpl$ and $n_s'=0.0023$. 
We also consider the case of $n_s=0.96$ and $n_s'=0.005$, which requires $A=3.1$, $B_0=3.08$ and $\alpha=0.01$
and has an inflationary energy scale of $V_0^{1/4}=0.004\mpl$. 
The latter
are the parameters analysed in Ref.~\cite{Bugaev:2009zh} 
and serve as a tool of comparison between our work and theirs. 
We also evaluate the spectra for the range of parameters which maximise the spectrum
in the range $k=\{10^{10},10^{14}\}[\rm{Mpc^{-1}}]$. This is because PBHs forming within this range
are possible canditates for Dark Matter. The parameters maximising the spectrum at $k=10^{10}[\rm{Mpc^{-1}}]$
are $A=4.4$, $B=4.2$, $\alpha=0.01$, and they satisfy $n_s=0.96$ and $n_s'=0.012$ with an inflationary energy scale of
$V_0^{1/4}=0.0011\mpl$. Maximising the spectrum at $k=10^{14}[\rm{Mpc^{-1}}]$ requires $A=3.4$, $B=0.01$ and $\alpha=0.01$,
satisfying $n_s=0.96$, $n_s'=0.0067$ with an inflationary energy scale of $V_0^{1/4}=0.0008\mpl$.
We also investigated whether this model could maximise the spectrum near $k=10^{5}[\rm{Mpc^{-1}}]$ since
these PBHs would be candidates for seeds of SuperMassive Black Holes \cite{Frampton:2010sw, Kawasaki:2012kn}, however we found that this would
require a running greater than that allowed by WMAP.
The
primordial scalar spectra for these choices are plotted in \fig{spectra_rmm}.

\section{Results and Discussion}\label{sec:results}

In both cases we begin by computing the integral over $y$ in \eq{eq:Ph} for a range of $k=10^4\cdots10^{25}\mathrm{Mpc}^{-1}$.
This is done using the known results of \eq{tauint} for a radiation era, \eqs{t1}{t2}, and the \emph{Second Euler-
Maclaurin summation formula} \cite{num_rec}. This method allows us to
avoid integrating over a singularity which occurs when $v=1$ by aiding us in `avoiding'
the endpoints. We introduce a sudden cutoff approximation for the calculation of the induced
gravitational wave spectrum, effectively we assume that structure on very small
scales $k>k_{end}$ is non-existent, $\calp_\zeta(k>k_{end})\sim0$. To ensure that
the PBH bound is not violated, we perform
a comparative analysis calculation, in that we set $\calp_\zeta=0$ for the scales
on which the model predicts a spectrum greater than the PBH bound. We have
engineered  the models so that after $N$ $e-$folds of inflation, our spectrum is just below
the PBH bound and that no more perturbations are produced afterwards.
In the case of the hilltop-model, we calculate the induced gravitational wave 
spectrum for $N=55$, $N=60$, and $N=65$, which we have chosen to reflect the standard
choices that appear in the literature, with the lower values of $N$ indicating a lower reheat temperature;
an issue which we are investigating in a follow up paper. In the case of the running mass model,
we select model parameters which both satisfy $n_s=0.96$ and $n_s'=0.002$ and $n_s'=0.005$ respectively
as well as the model parameters predicting the production of PBHs within the Dark Matter range, corresponding
to $0.0067<n_s'<0.012$.
Unlike in the hilltop model, no demand for the maximisation of the spectrum at a particular $e-$fold is made. 
Instead, if the spectrum touches the PBH bound, we terminate inflation, as can be seen in \fig{spectra_rmm}.

\begin{figure}
 \centering\includegraphics[width=4in, totalheight=4in]{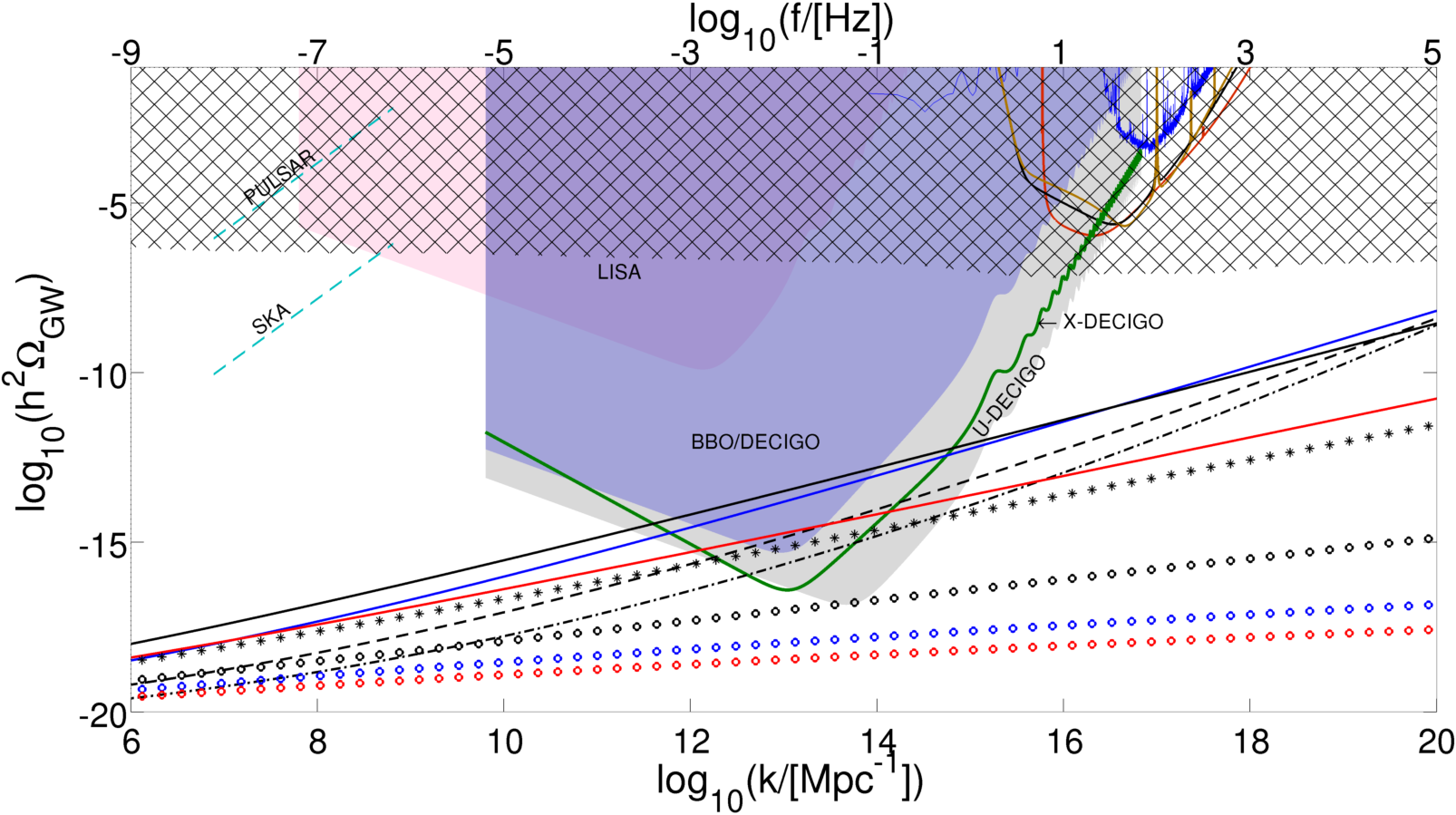}
\caption{Energy densities, \eq{eq:omega}, of the induced gravitational wave spectra from hilltop inflation 
with self-coupling powers labelled in the legend 
next to \fig{omega_hill3}. The plot corresponds to maximising the first order
spectrum at $N=55$.  The shaded regions correspond to
the regions of sensitivity of the gravitational wave detectors, with U-DECIGO corresponding
to the Ultimate DECIGO detector. The dashed blue lines  correspond to the sensitivity from the pulsar timing array. 
The thick solid green line corresponds to the cross-correlated DECIGO detector (X-DECIGO). The
thin straight lines in the upper right hand corner of the plots correspond to Advanced LIGO (red), the 6th run of LIGO (blue), the LCGT
 official data (black) and the LCGT updated data (brown). The cross-hatched region is the PBH constraint on the induced gravitational waves from Inflation.}
\label{omega_hill1}
\end{figure}

We plot the results of \eq{eq:Omega} for the induced gravitational waves from the hilltop model in \figs{omega_hill1}{omega_hill2}
and for the running mass model in \fig{omega_rmm}. We also plot the sensitivity curves from the various 
gravitational wave detectors, LIGO \cite{ligo}, LCGT \cite{lcgt}, LISA\cite{lisa} \footnote{We use the detector
parameters of the original NASA/ESA experiment which are available, however this is now a European only experiment \cite{AmaroSeoane:2012km}
and the detector specificiations have changed slightly, as dicussed in Appendix~(\ref{app:sens_curves}).}, BBO/DECIGO \cite{Seto:2001qf}, 
cross-correlated DECIGO and Ultimate DECIGO
as well as rough estimates of the current pulsar timing limit and the expected limit from the Square Kilometre Array
(SKA) \cite{Jenet, Hobbs, Yardley, PPTA}. 
We highlight the fact that our estimate for the cross-correlated DECIGO is rather basic by plotting
it differently to BBO/DECIGO and ultimate DECIGO. How these curves are generated and the data sets used are explained 
in Appendix~\ref{app:sens_curves}. The primordial gravitational wave spectrum can also be
large enough to be detectable by BBO and DECIGO, as was shown for single large field models of inflation
 (for example see Refs.~\cite{Seto:2001qf,Boyle:2005ug,Chongchitnan:2006pe,Kuroyanagi:2010mm}). 
However, this fact does not affect our conclusions since priomordial gravitational waves predict a scale invariant
spectrum, and has been shown here and in previous works (see for example Refs.~\cite{Ananda:2006af, Baumann:2007zm}) and in this paper, the spectrum
of induced gravitational waves is not as simple, and hence are distinguishable from each other.

\begin{figure}
\centering\includegraphics[width=4in,totalheight=4in]{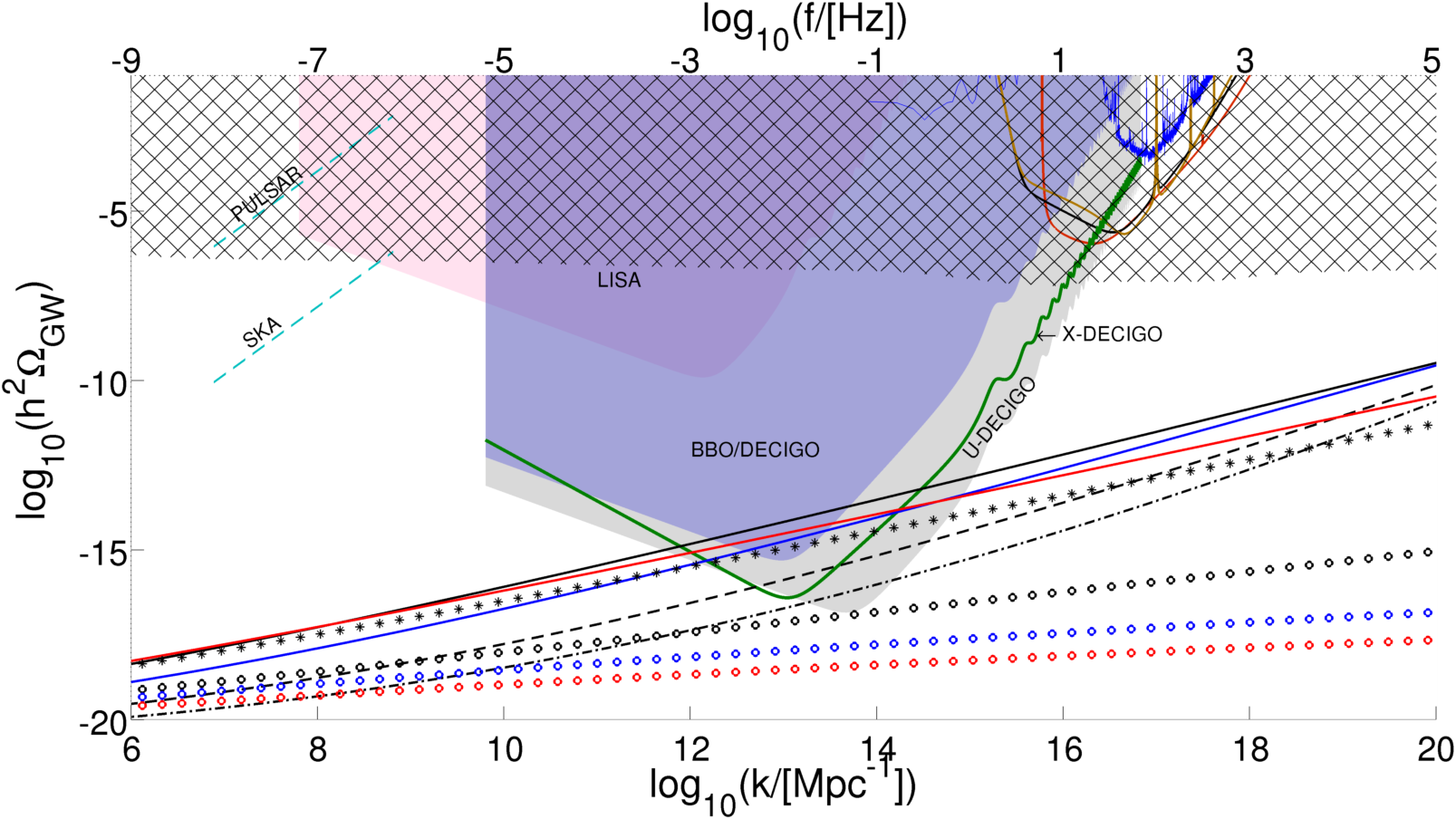}
\caption{ Energy densities, \eq{eq:omega}, of the induced gravitational wave spectra from hilltop inflation 
with self-coupling powers labelled in the legend 
next to \fig{omega_hill3}. The  plot corresponds to maximising the first order
spectrum at $N=60$.   }
\label{omega_hill2}
\end{figure}
\begin{figure}
\centering \centering\includegraphics[width=4in,totalheight=4in]{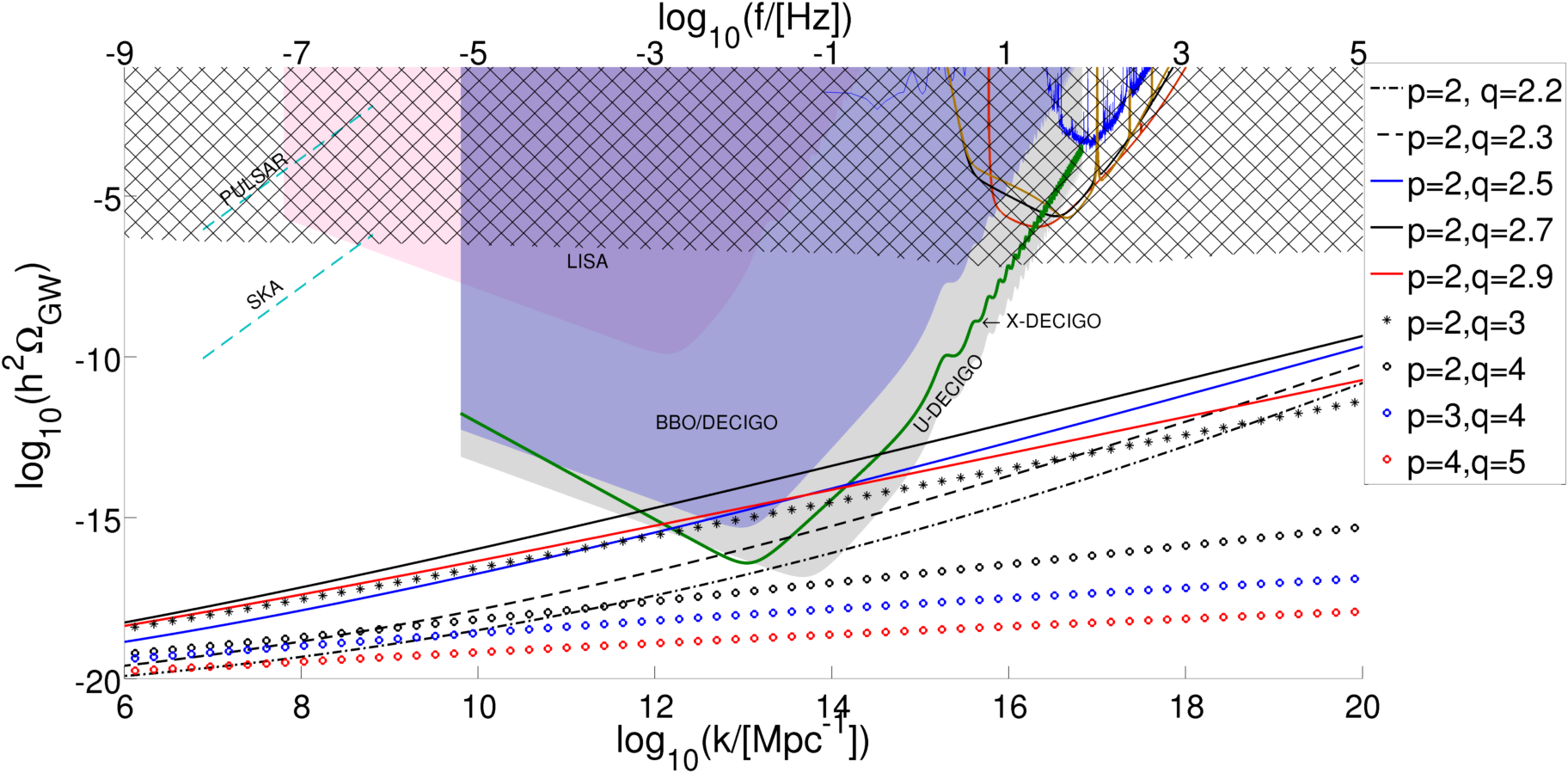}
\caption{The induced gravitational wave spectrum, \eq{eq:omega}, for the hilltop model with a scalar spectrum maximised at $N=65$ $e-$folds. The unlabelled lines in the figure are defined in \fig{omega_hill1}}
\label{omega_hill3}
\end{figure}

\begin{figure}
 \centering\includegraphics[width=4in,totalheight=4in]{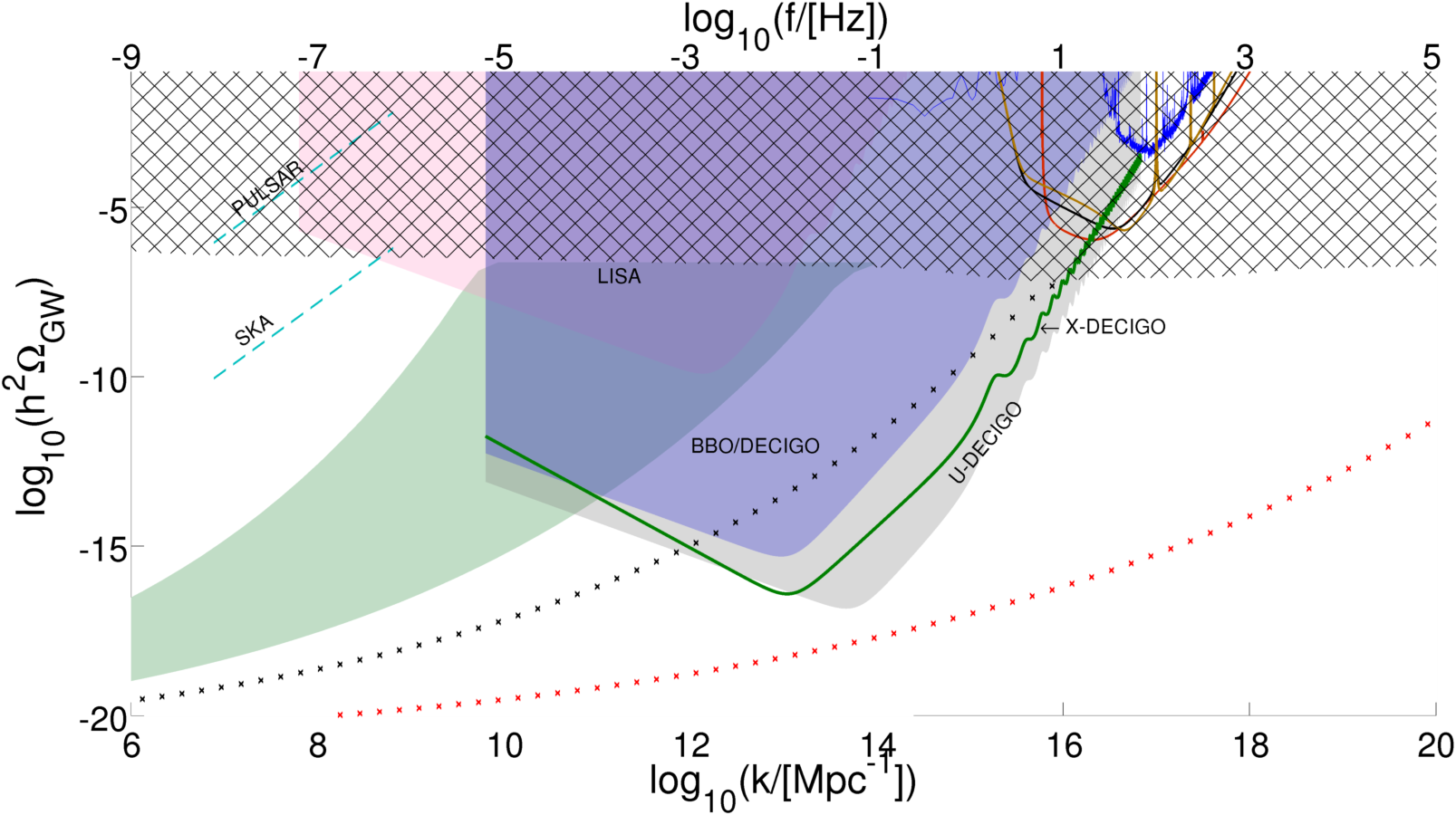}
\caption{ Running mass model predictions for the induced gravitational wave spectra, \eq{eq:omega}. 
The $n_s'=0.002$ (red crosses) model is evolved to $N\sim64$ 
while the $n_s'=0.005$ (black crosses) model requires that inflation terminates at 
 $N\sim43$ to satisfy the PBH bound. The green shaded
region corresponds to the range of parameters in which
result in the production
of PBHs whose energy density agrees with that of Dark Matter, and inflation
is required to terminate between $29\lesssim{}N\lesssim39$ to satisfy the PBH bound. 
The hatched region and upper right hand corner lines are all defined
in \fig{omega_hill3}.}
\label{omega_rmm}
\end{figure}

We find that the hilltop model with integral self coupling powers
$p=2$ and $q=3$ generates a GW spectrum detectable by the BBO/DECIGO
experiment for inflation lasting a reasonable number of
$e-$folds. Furthermore,  we find that to be detectable by the
BBO/DECIGO experiment, for inflation terminating within $55$
$e-$folds, the hilltop model requires coupling powers of $p=2$ and
$2.3\lesssim{}q\leq{}3$ while for $N=60$ the range $q$ is reduced to
$2.5\lesssim{}q\leq3$. In the former case $q=2.2$ comes within range
of the cross-correlated DECIGO and for the latter $q=2.3$ comes within
its range. It can be clearly seen in \figss{omega_hill1}{omega_hill2}{omega_hill3} that within the sensitivity
range of gravitational wave detectors, the results for the $N=60$
and the $N=65$ cases are the same, as expected. 

For the running mass model the $n_s'=0.005$ model requires
that inflation terminates after $43$ $e-$folds of inflation, which is
rather difficult to motivate, but predicts an induced gravitational
wave signal well within the range of BBO/DECIGO. The parameters which
lead to PBH candidates for Dark Matter come within the range of LISA
as well as BBO/DECIGO, but they also require an early termination
of inflation, between $29\lesssim{}N\lesssim39$. Since these parameter
choices
predict $V_0$ to be on the GUT scale, we can only reduce $N$ by
assuming a matter dominated phase of reheating and a reheat
temperature of $T_{RH}\sim1\rm{MeV}$.  However, this can be
problematic for the induced gravitational wave prediction, since the
scales of interest are so small, they re-enter the horizon immediately
after the end of inflation. As we mention in the introduction, and
several times thereafter, we assume a radiation dominated universe
during the formation of these gravitational waves.  Including an early
matter dominated phase would affect the predictions for the smallest
of scales,  making our results for the running mass model at the
largest $k$ values for $n_s'=0.005$, as well as the other parameter
values which fall within the green shaded region of \fig{omega_rmm},
questionable. On the other hand $N=65$ is an acceptable value and is compatible
with instant reheating into a radiation dominated universe. Unfortunately, the
spectra for induced gravitational waves for running mass models terminating between
$N=51$ and $N=65$ $e-$ folds do not seem to be within the sensitivity
range of future GW detectors, despite the fact that they do  result in
the formation of PBHs. It may be %
necessary to asses the impact a
matter dominated reheating phase \footnote{One can assume
other equations of state for this epoch, for example see Refs.~\cite{Joyce:1996cp, Seto:2003kc,Boyle:2007zx, Hidalgo:2011fj}.}
 could have on the eventual induced
spectrum of gravitational
waves~\cite{Seto:2003kc,Boyle:2007zx,Nakayama:2008ip,Nakayama:2008wy}.

We should note here that Ref.~\cite{Chluba} have recently used the COBE-FIRAS data to further constrain the spectrum on $k<10^4{\rm Mpc}^{-1}$. They also conclude that the upcoming PIXIE experiment
will further constrain this small scale spectrum, which in turn could rule out the running mass
model.

\section{Conclusions}\label{sec:conclusion}

We have found that for a reasonable range of $e-$folds,
Hilltop-type models predict  a spectrum of induced gravitational waves
likely to be detectable by both the DECIGO and cross correlated
DECIGO. More interestingly, the model with integral coupling powers
$p=2,q=3$, which is strongly motivated in particle physics models,  is
within the range of the detection, while satisfying the PBH bounds and
WMAP constraints. This is shown in \figss{omega_hill1}{omega_hill2}{omega_hill3}.

On the other hand, the running mass model may also lead to a spectrum
within the range of the LISA and DECIGO experiment, if small values of $N$  can
be motivated. In which case, the produced PBHs, with masses within
the range $M_{\rm BH} \sim
10^{20} -10^{27}$~g, can be candidates for dark matter. 
This scenario will be checked by
the future gravitational wave observations (See Fig.\ref{omega_rmm}). 

We hope the earliest possible completion of these new types of
gravitational wave observatories.

\acknowledgments
We thank  A. Christopherson, 
D. Galliano, T. Hiramatsu, D.H. Lyth, R. Saito,  A. Taruya, T. Tanaka, and J. Yokoyama for useful
discussions. This work was supported in part by grant-in-aid from the
Ministry of Education, Culture, Sports, Science, and Technology (MEXT)
of Japan,  No. 2200775 (L.A.), No. 21111006, No. 23540327, No.22244030
(K.K.). L.A. is also supported by the Japanese Society for the Promotion of Science
(JSPS). M.S. acknowledges Monbukagaku-sho Grant-in-Aid for the Global COE programs,
”The Next Generation of Physics, Spun from Universality
and Emergence” at Kyoto University, and
JSPS Grant-in-Aid for Scientific Research (A) No.~21244033. K.K. was partly supported by the
Center for the Promotion of Integrated Sciences (CPIS) of Sokendai,
No.~1HB5806020.

\appendix

\section{Induced Gravitational Waves}\label{app:PGW}
In this section we will briefly review the equations relevant to the formation
and evolution of gravitational waves sourced by the primordial scalar perturbations 
\cite{Mollerach:2003nq,Ananda:2006af, Baumann:2007zm}.
These induced gravitational waves are in essence a physical manifestation of the spatial
perturbations which arise from
taking the Taylor expansion of the 
metric up to order second. The ADM metric in this case is given as

\begin{equation}\label{ADM2}
ds^2=a^2(\tau)\left[-\left(1+2\Phi^{(1)}+2\Phi^{(2)}\right)d\tau^2+2V_i^{(2)}d\tau{}dx^i+
\left\{\left(1-2\Phi^{(1)}-2\Phi^{(2)}\right)\delta_{ij}+\frac{1}{2}h_{ij}\right\}dx^idx^j\right]
\end{equation}
where we have chosen a longitudinal gauge and assume $\Phi=\Psi$ at all orders, $\Phi$ is the Bardeen potential,
first order vector perturbations are ignored ($V^{(1)}=0$) 
and the tensor perturbations
($h_{ij}$) include both first and second order effects. 
In this paper we
neglect anisotropic stress.  The next step is to calculate the Einstein equations\cite{Acquaviva:2002ud,Baumann:2007zm}:

\bea
G^{(2)i}_j&=&a^{-2}\left[\frac{1}{4}\left(h^{i''}_j+2\mathcal{H}h^{i'}_j-\nabla^2h^i_j\right)+4\Phi^{(1)}\partial^i\partial_j\Phi^{(1)}
+2\partial^i\Phi^{(1)}\partial_j\Phi^{(1)}\right.\nonumber\\
&&\left.+[\rm{Second~Order~Terms}]+[\rm{diagonal~terms}]\right]
\label{G2}
\eea
where the $\rm{Second Order Terms}$ refer to terms containing second order scalar perturbations. The spatial part of the energy-momentum tensor is:
\be
T^{(2)i}_j=(\rho^{(0)}+P^{(0)})v^{(1)i}v^{(1)}_j+P^{(2)}\delta^i_j
\label{T2}
\ee
where $\rho$ and $P$ are the energy density and pressure, and $v$ is the velocity. Here on, for simplicity, our notation
is such that $h_{ij}=h^{(2)}_{ij}$.

As is standard, to evaluate the spectrum of these gravitational waves the Fourier mode of the tensor perturbation is taken:
\be
h_{ij}(\mathbf{x},\tau)=
\frac{1}{(2\pi)^{3/2}}\int{}
d^3k{}e^{i\mathbf{k}\cdot\mathbf{x}}
\left[h_\mathbf{k}(\tau)\rme_{ij}(\mathbf{k})+\bar{h}_\mathbf{k}(\tau)\bar{\rme}_{ij}(\mathbf{k})\right]
\label{h2fourier}
\ee
where the polarization tensors $\rme,\bar{\rme}$ are given in terms of the orthonormal vectors $(\mathbf{e},
\bar{\mathbf{e}})$ \cite{Baskaran:2006qs, Ananda:2006af,Baumann:2007zm}:

\bea\label{Pol}
\rme_{ij}(\mathbf{k})=\frac{1}{\sqrt{2}}\left[e_{i}e_j-\bar{e}_i\bar{e}_j\right]&~~~&
\bar{\rme}_{ij}(\mathbf{k})=\frac{1}{\sqrt{2}}\left[e_{i}\bar{e}_j+\bar{e}_{i}e_j\right]
\eea
these vectors have been defined to satisfy the conditions that (a) the
gravitational waves are traceless, thus $\rme_{ij}\delta^{ij}=0$, (b)normal and (c) transverse .

According to Refs.~\cite{Ananda:2006af, Baumann:2007zm}, there exists a projection tensor defined robustly in Ref.~\cite{Ananda:2006af}, 
which extracts the transverse, traceless
parts of \eq{G2} and also does away with the $\rm{Second~Order~Terms}$:
\be\label{projection}
\hat{\mathcal{T}}^{lm}_{ij}G^{(2)}_{lm}=8\pi{}G\hat{\mathcal{T}}^{lm}_{ij}T^{(2)}_{lm}~.
\ee
Thus the equation of motion for the tensor perturbation can be derived,  using the following definitions \cite{Baumann:2007zm}:
\bea
P^{(0)}=w\rho^{(0)}\label{P0} &~~~&
\rho^{(0)}=\frac{3\mathcal{H}^2}{8\pi{}Ga^2}\label{rho0}\nonumber\\
v^{(1)}_i&=&-\frac{2}{8\pi{}Ga^2(1+w)\rho^{(0)}}\partial_i(\Phi'+\mathcal{H}\Phi)\label{v1}\,
\eea
$w$ is known as
the equation of state, $G$ is the Gravitational constant and we have dropped the superscript $(i)$ from
the Bardeen potential and from now on $\Phi$ refers to the first-order Bardeen potential.
Therefore from \eqss{G2}{T2}{projection}, and after some tidying
up one gets: 

\be\label{heq}
h''_{ij}+2\mathcal{H}h'_{ij}-\nabla^2h_{ij}=-4\hat{\mathcal{T}}^{lm}_{ij}\mathcal{S}_{lm}
\ee
with the source term:
\be\label{source}
\mathcal{S}_{ij}=4\Phi\partial_i\partial_j\Phi+2\partial_i\Phi\partial_j\Phi-\frac{4}{3\mathcal{H}^2(1+w)}\partial_i(\Phi'+\mathcal{H}\Phi)\partial_j(\Phi'+\mathcal{H}\Phi)~.
\ee

To move to Fourier space, the Fourier transform of the scalar quantity $\Phi$ is written down:
\be\label{PHIF}
\Phi(\mathbf{x})=\frac{1}{(2\pi)^{3/2}}\int{}d^3k{}e^{i\mathbf{k}\cdot\mathbf{x}}\Phi_\mathbf{k}
\ee
terms like $\partial_l\Phi$ then will pull down a $ik_l$ term and $\partial_l\partial_m\Phi$ will 
pull down a $-k_lk_m$ term, which from \eqs{source}{PHIF} leads to terms of the form:
\bea
\Phi\partial_l\partial_m\Phi&=&-\frac{1}{(2\pi)^3}\int{}d^3ke^{i\mathbf{k}\cdot\mathbf{x}}\left[\int{}d^3qq_lq_m\Phi_{\mathbf{k}-\mathbf{q}}\Phi_\mathbf{q}\right]\nonumber\\
\partial_l\Phi\partial_m\Phi&=&-\frac{1}{(2\pi)^3}\int{}d^3ke^{i\mathbf{k}\cdot\mathbf{x}}\left[\int{}d^3q(k_l-q_l)k_m\Phi_{\mathbf{k}-\mathbf{q}}\Phi_\mathbf{q}\right]
\eea
The Fourier form of \eq{heq} is then:

\be
h''_\mathbf{k}\rme_{ij}+\bar{h}''_\mathbf{k}\bar{\rme}_{ij}+2\calh{}h'_\mathbf{k}\rme_{ij}
+2\calh\bar{h}'_\mathbf{k}\bar{\rme}_{ij}+k^2h_\mathbf{k}\rme_{ij}+k^2\bar{h}_\mathbf{k}\bar{\rme}_{ij}=\cdots
\ee
where the $\cdots$ refer to the Fourier transform of the right hand side of \eq{heq}
and $h_\mathbf{k}=h_\mathbf{k}(\tau)$. To extract 
$h$ and get rid of $\bar{h}$ one simply multiplies through by $\rme^{ij}(\mathbf{k})$ to get:

\bea\label{eqf}
h''_\mathbf{k}+2\calh{}h'_\mathbf{k}+k^2h_\mathbf{k}&=&
4\int{}\frac{d^3{q}}{(2\pi)^{3/2}}\rme^{ij}(\mathbf{k})\frac{q_iq_j}{3(1+w)}\Large[\nonumber\\
&&(10+6w)\Phi_\mathbf{q}\Phi_{\mathbf{k}-\mathbf{q}}+\frac{8}{\calh}\Phi_\mathbf{q}\Phi'_{\mathbf{k}-\mathbf{q}}
+\frac{4}{\calh^2}\Phi'_\mathbf{q}\Phi'_{\mathbf{k}-\mathbf{q}}\Large]~.
\eea
To solve this equation, the usual change of variables $ah_\mathbf{k}=v_\mathbf{k}$ is made, and \eq{eqf} 
becomes:

\be\label{veq}
v''_\mathbf{k}+\(k^2-\frac{a''}{a}\)v_\mathbf{k}=a\mathcal{S}\,
\ee
where $\mathcal{S}$ is defined via \eq{eqf} and is written down
explicitly as
\be
a\mathcal{S} \equiv 
4\int{}\frac{d^3{q}}{(2\pi)^{3/2}}\rme^{ij}(\mathbf{k})\frac{q_iq_j}{3(1+w)}\left[
(10+6w)\Phi_\mathbf{q}\Phi_{\mathbf{k}-\mathbf{q}}+\frac{8}{\calh}\Phi_\mathbf{q}\Phi'_{\mathbf{k}-\mathbf{q}}
+\frac{4}{\calh^2}\Phi'_\mathbf{q}\Phi'_{\mathbf{k}-\mathbf{q}}\right]~.
\ee
The solution of \eq{veq} is found via the Greens' function method
i.e the solution is written down as:
\be\label{hsol}
h_\mathbf{k}(\tau)=\frac{1}{a(\tau)}\int{}d\tilde{\tau}g_\mathbf{k}(\tau;\tilde{\tau})[a(\tilde{\tau})\mathcal{S}(\mathbf{k},\tilde{\tau})]
\ee
the Greens function is then the solution of:
\be\label{greeneq}
g''_\mathbf{k}+\(k^2-\frac{a''}{a}\)g_\mathbf{k}=\delta(\tau-\tilde{\tau})~.
\ee
To solve \eq{greeneq} the functional form of the scale factor is needed, and hence the epoch should be defined.
Therefore $g$ has two forms, one for matter domination (MD) and another for radiation domination (RD). The derivation
of which can be found in Appendix A of Ref.~\cite{Baumann:2007zm}, here we just state the results:

\bea
g_k(\tau;\tilde{\tau})&=&\frac{1}{k}\sin\left[k(\tau-\tilde{\tau})\right]\label{RD}\\
g_k(\tau;\tilde{\tau})&=&-\frac{x\tilde{x}}{k}\left[j_1(x)y_1(\tilde{x})-j_1(\tilde{x})y_1(x)\right]\label{MD}
\eea
where \eq{RD} is for RD, \eq{MD} is for MD, $x=k\tau$ and $j_1,y_1$ are the spherical Bessel functions. 

In order to evaluate the source term, the expressions for the Bardeen potential are also required. 
Since it is a first order quantity it is obtained from linear theory \cite{Dodelson:2003ft}, 
\be
\Phi''_\mathbf{k}+\frac{6(1+w)}{1+3w}\frac{1}{\tau}\Phi'_\mathbf{k}+wk^2\Phi_\mathbf{k}=0
\ee
and can be solved exactly using Bessel functions:
\be
\Phi_\mathbf{k}(\tau)=\tilde{y}^{-\alpha}\left[C_1(k)J_\alpha(\tilde{y})+C_2(k)Y_\alpha(\tilde{y})\right]
\ee
where $w>0$ and 
\bea
\tilde{y}=\sqrt{w}k\tau&~~~~&\alpha=\frac{1}{2}\left(\frac{5+3w}{1+3w}\right)~.
\eea

For Matter Domination $w=0$ and:

\be
\Phi_\mathbf{k}(\tau)=C_1(k)\,
\ee
where we drop the decaying mode, and during Radiation Domination
\be
\Phi_\mathbf{k}(\tau)=\frac{1}{\tilde{y}^2}C_1(k)\(\frac{\sin(\tilde{y})}{\tilde{y}}-\cos(\tilde{y})\)~.
\ee
In the limit of small $\tilde{y}$ (early times) this function reduces to $C_1(k)$ and 
at early times it should be equal to the vacuum perturbation  $\psi_\mathbf{k}$,
$\Phi_\mathbf{k}(\tau\to0)\to\psi_\mathbf{k}$. Then we can write:

\be
\Phi_\mathbf{k}(\tau)=\psi_\mathbf{k}\Phi(k\tau)
\ee
where the $\Phi$ on the right hand side is the transfer function
which evolves the primordial fluctuation $\psi_\mathbf{k}$, 
and we hope the notational degeneracy is acceptable to the reader.%

The main aim is to calculate the spectrum of induced gravitational waves,
which
is essentially a measure of the correlation between two modes:

\be
<h_\mathbf{k}h_\mathbf{k'}>=\frac{2\pi^2}{k^3}\delta(\mathbf{k}+\mathbf{k}')P_h(k,\tau)\,
\ee
and all the ingredients needed to evaluate this spectrum are now in place. The spectrum of \eq{hsol} is
\be
<h_\mathbf{k}h_\mathbf{k'}>=\frac{1}{a^2}\int_0^{\tau}d\tau_1d\tau_2g_k(\tau;\tau_1)
g_k(\tau;\tau_2)a(\tau_1)a(\tau_2)<\mathcal{S}(\tau_1,\mathbf{k})\mathcal{S}(\tau_2,\mathbf{k})>~.
\ee
and in this work we are only interested in the RD part. 
Before we piece things together, we make a small
redefinition: $\rme(\mathbf{k},\mathbf{q})
=\rme^{ij}(\mathbf{k})q_iq_j=q^2(1-\mu^2)$ where $\mu=(\mathbf{k}\cdot\mathbf{q})/(kq)$.

The source term can be written as:
\be
\mathcal{S}(\mathbf{k},\tau)=\frac{1}{(2\pi)^{3/2}}\int{}d^3{q}\rme(\mathbf{k},\mathbf{q})f(\mathbf{k},\mathbf{q},\tau)\psi_\mathbf{q}\psi_{\mathbf{k}-\mathbf{q}}
\ee
where
\be
f(\mathbf{k},\mathbf{q},\tau)=\frac{2(5+3w)}{3(1+w)}\Phi_\mathbf{q}\Phi_{\mathbf{k}-\mathbf{q}}+
\frac{8}{3\calh(1+w)}\Phi_\mathbf{q}\Phi'_{\mathbf{k}-\mathbf{q}}+\frac{4}{3\calh^2(1+w)}\Phi'_\mathbf{q}\Phi'_{\mathbf{k}-\mathbf{q}}~.
\ee
The primordial spectrum of linear scalar perturbations is given by:
\be
<\psi_\mathbf{q}\psi_\mathbf{k}>=\frac{2\pi^2}{q^3}P(q)\delta(\mathbf{q}+\mathbf{k}).
\ee
Thus the correlation between two modes of the source term can finally be written as:
\bea
<\mathcal{S}(\mathbf{k},\tau_1)\mathcal{S}(\mathbf{k'},\tau_2)>&=&\delta(\mathbf{k}+\mathbf{k'})\int{}d^3\tilde{{k}}\rme(\mathbf{k},\tilde{\mathbf{k}})^2f(\mathbf{k},\tilde{\mathbf{k}},\tau_1)
\left[f(\mathbf{k},\tilde{\mathbf{k}},\tau_2)\right.\nonumber\\
&&\left.+f(\mathbf{k},\mathbf{k}-\tilde{\mathbf{k}},\tau_2)\right]\frac{P(|\mathbf{k}-\tilde{\mathbf{k}}|)}{|\mathbf{k}-\tilde{\mathbf{k}}|^3}\frac{P(\tilde{k})}{\tilde{k}^3}
\eea
and the spectrum of induced gravitational waves is:
\be
P_h(k)=\frac{1}{a^2}\int_0^\infty{}d\tilde{k}\int_{-1}^{1}d\mu\frac{k^3\tilde{k}^3}{|\mathbf{k}-\tilde{\mathbf{k}}|^3}(1-\mu^2)^2P(|\mathbf{k}-\tilde{\mathbf{k}}|)P(\tilde{k})I_1(k,\tilde{k},\tau)I_2(k,\tilde{k},\tau)
\ee
where
\bea
I_1(k,\tilde{k},\tau)&=&\int{}d\tau_1a(\tau_1)g_k(\tau;\tau_1)f(k,\tilde{k},\tau_1)\nonumber\\
I_2(k,\tilde{k},\tau)&=&\int{}d\tau_2a(\tau_2)g_k(\tau;\tau_2)[f(k,\tilde{k},\tau_2)+f(k,|\mathbf{k}-\tilde{\mathbf{k}}|,\tau_2)]
\eea
are generic integrals, and depend only on the epoch of evaluation. We present their results in the next
section for radiation domination. First we rewrite the spectrum in terms of the variables $v=\tilde{k}/k$, 
$y=\sqrt{1+v^2-2v\mu}$, and $x=k\tau$. During RD $a\propto\tau$ and the spectrum reduces to

\be
P_h(k)=\frac{k^2}{x^2}\int_0^\infty{}dv\int_{|v-1|}^{|v+1|}dy\frac{v^2}{y^2}(1-\mu^2)^2P(kv)P(ky)\tilde{I}_1\tilde{I}_2
\label{Pv}
\ee
where
\bea\label{tauint}
I_1(k,\tilde{k},\tau)&=&\frac{1}{k^2}\tilde{I}_1\nonumber\\
&=&\frac{1}{k^2}\int{}dx_1x_1g_k(x;x_1)f(k,x_1v)\nonumber\\
I_2(k,\tilde{k},\tau)&=&\frac{1}{k^2}\tilde{I}_2\nonumber\\
&=&\frac{1}{k^2}\int{}dx_2x_2g_k(x;x_2)[f(k,x_2v)+f(k,x_2y)]
\eea
\subsection{The $\tau$ integrals}
This was first shown in Ref.~\cite{Ananda:2006af}. The first integral
in \eq{tauint} is found to be: \bea\label{t1}
\tilde{I}_1&=&\frac{1}{4ky^3v^3}\left\{-\cos(x)\sum_{n=1}^{4}\alpha_n\rm{Si}(\beta_nx)
  +\sin(x)\sum_{n=1}^{4}(-1)^{n+1}\alpha_n\rm{ci}(\beta_nx)\right\}\nonumber\\
&&+\gamma_1\sin(x)+\gamma_2\sin(vx)\sin(yx)+\gamma_3\sin(vx)\cos(yx)+
\gamma_4\cos(vx)\sin(yx)+\gamma_5\cos(vx)\cos(yx)\nonumber\\
&& \eea and the second integral is given by: \bea\label{t2}
\tilde{I}_2&=&-\frac{\alpha}{2kv^3y^3}\left\{\cos(x)\left[-\rm{Si}(\beta_1x)+\rm{Si}(\beta_2x)+\rm{Si}(\beta_3x)-\rm{Si}(\beta_4x)
  \right]\right.\nonumber\\
&&\left.+\sin(x)\left[\rm{ci}(\beta_1x)+\rm{ci}(\beta_2x)-\rm{ci}(\beta_3x)-\rm{ci}(\beta_4x)\right]\right\}\nonumber\\
&&+\gamma_{21}\sin(x)+\gamma_{22}\sin(x(v+y))+
\gamma_{23}\sin(x(v-y))+\gamma_{24}\cos(x(v-y))+\gamma_{25}\cos(x(v+y))\nonumber\\
&& \eea 
the coefficients in these two integrals are given in Appendix \ref{app:tau}.  The $\alpha_n$ coefficients in
the first integral are found to have the property that
$\sum_{n=1}^{4}(-1)^n\alpha_n=0$  and  the $\alpha$
coefficient in the second integral is given by $\alpha=(v^2-1+y^2)^2$.

\begin{figure}
\begin{minipage}{0.49\linewidth}
 \includegraphics[width=3 in, totalheight=3 in]{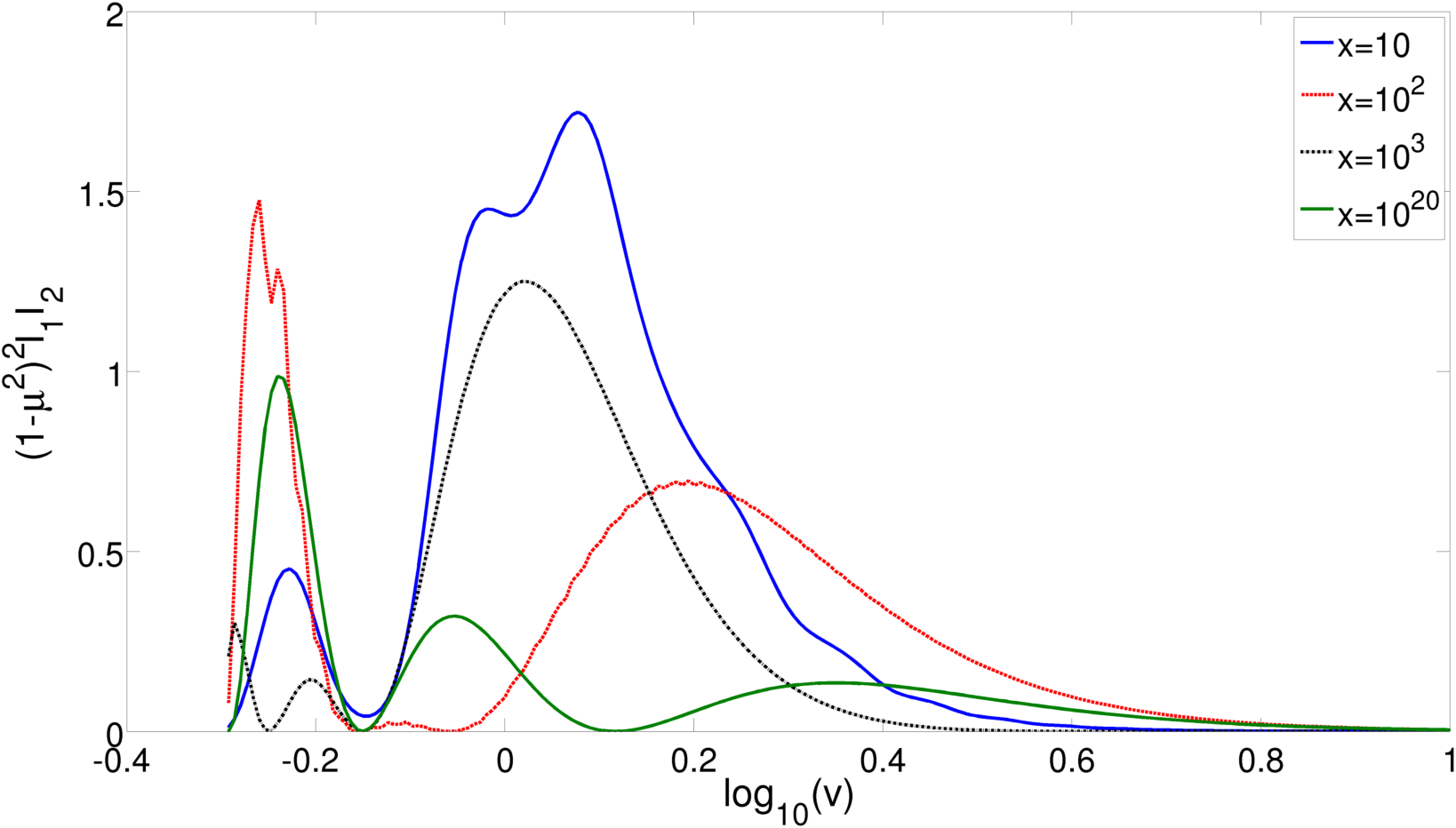}
\end{minipage}
\hspace{0.5cm}
\begin{minipage}{0.49\linewidth}
 \includegraphics[width=3 in, totalheight=3 in]{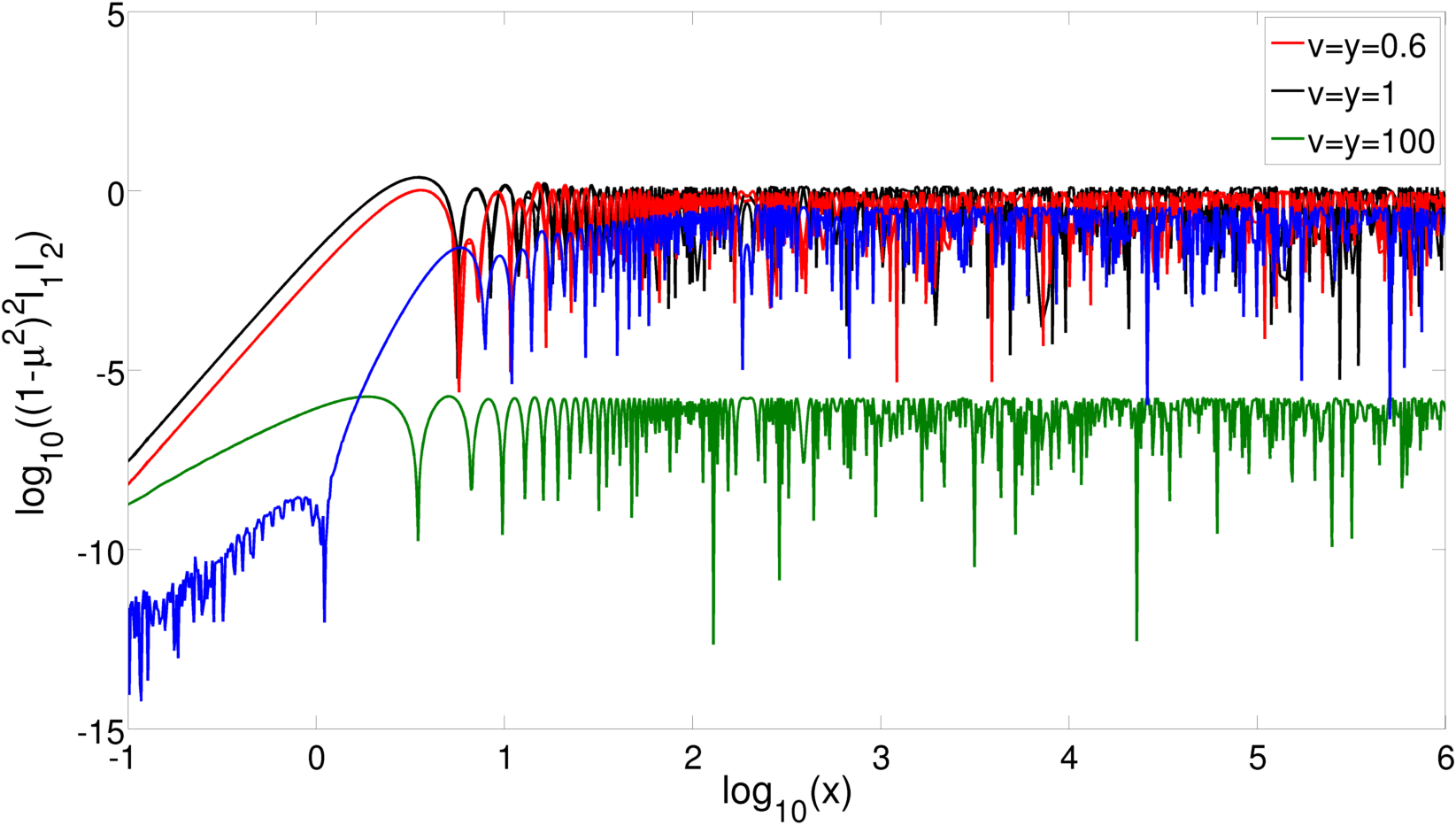}
\end{minipage}
\caption{\label{tau_int}Plot of the $\tau$ integrals for fixed values of $x=k\tau$ and varying $v$ (left), where we have set $y=v$, which clearly shows that the integrals 
asymptote to zero for large $v$ meaning
we need not integrate over an infinite range of scales. Plot of the $\tau$ integrals for fixed $v=y$ and varying $x=k\tau$ (right), this shows that (a) the effect
is physical since the
source term peaks after horizon entry ($x>1$) and (b) the envelope of the integrals rapidly asymptotes to a constant value for fixed
$v$ and $y$. 
}
\end{figure}

The $\rm{Si}$ and $\rm{ci}$ terms are the sine and cosine integrals respectively \cite{abr}, defined as:

\bea
\rm{Si}(x)&=&\int_0^x\frac{\sin(t)}{t}dt\,\nonumber\\
\rm{ci}(x)&=&\int_0^x\frac{\cos(t)}{t}dt~.
\eea
In the limit of large $x$ the sine integral asymptotes to a constant value $Si(x)\to(1+\gamma_{euler})$
and the cosine integral asymptotes to $ci(x)\to\ln(x/\gamma_{euler})$, where $\gamma_{euler}$ is the 
Euler-Mascheroni constant. The integrals are found to asymptote to $0$ for large $v$ and $y$, as well
as for when $v,y$ or $x$ approach $0$. These are useful properties, and it means that \eq{Pv} need not be evaluated
for $v\to\infty$ but to a much smaller value, and the final results will be the same. We plot the properties 
of these integrals in \fig{tau_int}, which shows agreement between our numerics and those of Refs.~\cite{Ananda:2006af} and \cite{Saito:2009jt}.

\subsection{The energy density of second order gravitational waves}\label{app:omega}
Detectors of gravitational waves will measure the amplitude of their energy
density which can be parametrized 
by the dimensionless variable (e.g. Ref.~\cite{Maggiore:2000gv}):

\be
\Omega_{\rm GW}(f)=\frac{1}{\rho_c}\frac{d\rho_{\rm GW}}{d\ln{}f}
\ee
where $f$ is the frequency, $\rho_c$ is the critical energy density defining
the coasting solution of the Friedman equation and $\rho_{\rm GW}$ is the energy
density of gravitational waves. This parameter is related to the spectrum of
induced gravitational waves $P_h$ as \cite{Baumann:2007zm}
\be\label{omega2}
\Omega_{\rm GW}^{(2)}=\frac{a(\tau)}{a_{\rm eq}k_{\rm eq}^2}t^2(k,\tau)P_h(k,\tau)\,
\ee
where $eq$ denotes matter-radiation equality, and $t$ is the transfer function
found to be \cite{Baumann:2007zm}
\be
t(k,\tau)=\frac{a_{\rm eq}k_{\rm eq}}{a(\tau)k}
\ee
for scales $k\gg{}k_{\rm eq}$, our regime of interest and $k_{\rm eq}\simeq0.01\rm{Mpc}^{-1}$. This reduces
\eq{omega2} to:

\be
\Omega_{\rm GW}^{(2)}(k,z)=\frac{(1+z)}{(1+z_{\rm eq})}P_h(k)
\ee
where we made the $\tau\sim{}k^{-1}$ substitution, the time when a scale enters the horizon.
Thus today $\Omega_{\rm GW}(k,\eta_0)\simeq\frac{\calp_h(k)}{3300}$,
where $1+z_{\rm eq}\simeq3300$.

\section{The coefficients of the $\tau$ integral}\label{app:tau}
We present the coefficients of the time integrals (\ref{t1}) and (\ref{t2}) in the following tables.

\begin{table}[h]
\begin{center}
\begin{tabular}{|l|c|}
\hline
$\beta_1$&$1+v+y$\\
$\beta_2$&$-1+v+y$\\
$\beta_3$&$1+v-y$\\
$\beta_4$&$-1+v-y$\\
\hline
\end{tabular}
\begin{tabular}{|c|c|c|}
\hline
Coefficient&Symbol&Expression\\
\hline
$\sin(x)$&$\gamma_1$&$\frac{1}{kv^2y^2}(v^2-3y^2+1)$\\
$\sin(vx)\sin(yx)$&$\gamma_2$&$\frac{1}{kx^3y^3v^3}(2-x^2-x^2y^2+3x^2v^2)$\\
$\sin(vx)\cos(yx)$&$\gamma_3$&$-\frac{2}{ky^2x^2v^3}$\\
$\cos(vx)\sin(yx)$&$\gamma_4$&$-\frac{2}{kx^2y^3v^2}$\\
$\cos(vx)\cos(yx)$&$\gamma_5$&$\frac{2}{kxy^2v^2}$\\
\hline
$\sin(x)$&$\gamma_{21}$&$\frac{2}{kv^2y^2}(1-v^2-y^2)$\\
$\sin(x(v-y))$&$\gamma_{22}$&$\frac{2}{kx^2v^3y^3}(v-y)$\\
$\sin(x(v+y))$&$\gamma_{23}$&$-\frac{2}{kx^2v^3y^3}(v+y)$\\
$\cos(x(v-y))$&$\gamma_{24}$&$-\frac{1}{kx^3y^3v^3}(-2+x^2-x^2y^2-x^2v^2-2x^2vy)$\\
$\cos(x(v+y))$&$\gamma_{25}$&$-\frac{1}{kx^3v^3y^3}(2-x^2+x^2y^2+x^2v^2-2x^2vy)$\\
\hline
\end{tabular}
\end{center}
\caption{Table on the left gives the expressions for the coefficients of 
the arguments of the Cosine and Sine integrals in \eqs{t1}{t2}.
The right table gives the expressions for the $\tau_1$ integral \eq{t1} (tob block)
and for the $\tau_2$ integral \eq{t2} (bottom block).}
\label{beta}
\end{table}

\begin{table}
\centering
 \begin{tabular}{|l|c|c|c|c|c|c|c|}
 \hline
&$1$&$v^4$&$4v^3$&$4v^2$&$3y^4$&$4y^3$&$2y^2v^2$\\
\hline
$\alpha_1$&$-$&$+$&$+$&$+$&$-$&$-$&$-$\\
$\alpha_2$&$+$&$-$&$+$&$-$&$+$&$-$&$+$\\
$\alpha_3$&$+$&$-$&$-$&$-$&$+$&$-$&$+$\\
$\alpha_4$&$-$&$+$&$-$&$+$&$-$&$-$&$-$\\
\hline
\end{tabular}
 \caption{This table gives the expressions of the coefficients of the sine and cosine
integrals in \eq{t1}. Each $\alpha_n$ coefficient has the same parameters
as the others, but the parameters differ in their respective signs. 
The columns to the right of the $\alpha$s give the sign of the parameter defined
in the column header. For example then we can read off $\alpha_1$ as $-1+v^4+4v^3+4v^2-3y^4-4y^3-2y^2v^2$.}
\label{tab:alpha}
\end{table}

\section{The coefficients of the Hilltop Model}\label{app:hill_coeff}
In this section we list the values of $\eta_p$ and $\eta_q$ for the various 
coupling powers $p$ and $q$ in the hilltop model, \eq{hilltop-potential}. In each case
the model parameters are chosen to satisfy $n_s=0.95$, and the 
selection criteria is explained fully in Section~\ref{sec:hilltop}. We draw the readers attention
to the horizontal dashed line in the tables, this line separates the fractional powers 
from the integral powers, and it is clear that for the integral powers, the mass coupling
parameters are most strongly constrained by the bound on the spectral index and not from 
the maximisation of spectrum at the end of inflation.

\begin{table}[h]
 \begin{tabular}{|c|c|c|c|c|c|c|}
\hline
$p$&$q$&$\eta_p$&$\eta_q$&$\calp_\zeta(k_{end})$&$n_s'(k_0)$&$V_0^{1/4}$\\
 \hline
$2$&$2.2$&$0.17959$&$0.1449$&$0.017069$&$0.0059605$&$0.0014$\\
$2$&$2.3$&$0.13673$&$0.09898$&$0.014079$&$0.0078104$&$0.0015$\\
$2$&$2.5$&$0.1051$&$0.070408$&$0.01391$&$0.012493$&$0.0015$\\
$2$&$2.7$&$0.089583$&$0.045$&$0.0071942$&$0.016684$&$0.0017$\\
$2$&$2.9$&$0.071667$&$0.036667$&$0.00043916$&$0.016381$&$0.0015$\\
\hdashline
$2$&$3$&$0.065417$&$0.03125$&$0.00016705$&$0.016261$&$0.0015$\\
$2$&$4$&$0.038333$&$0.0083333$&$2.239e-06$&$0.016928$&$0.0015$\\
$3$&$4$&$0.041667$&$0.025$&$1.766e-07$&$0.016712$&$0.0012$\\
$4$&$5$&$0.0375$&$0.028421$&$7.4549e-08$&$0.016752$&$0.0009$\\

\hline
 \end{tabular}
\caption{Values of the coupling masses which satisfy the 
WMAP bounds at the pivot scale $k_0=0.002\rm{Mpc}^{-1}$ and maximise the 
spectrum at $N=55$ $e-$folds. Also included is energy scale of the model
evaluated at $k=k_0$ in units of $\rm{\mpl}$.}
\end{table}

\begin{table}[h]
 \begin{tabular}{|c|c|c|c|c|c|c|}
\hline
$p$&$q$&$\eta_p$&$\eta_q$&$\calp_\zeta(k_{end})$&$n_s'(k_0)$&$V_0^{1/4}$\\
\hline
$2$&$2.2$&$0.16939$&$0.15102$&$0.014497$&$0.0051655$&$0.0008$\\
$2$&$2.3$&$0.128$&$0.095833$&$0.014173$&$0.0066689$&$0.0013$\\
$2$&$2.5$&$0.098571$&$0.083333$&$0.01452$&$0.010865$&$0.0009$\\
$2$&$2.7$&$0.085417$&$0.058333$&$0.010977$&$0.015165$&$0.001$\\
$2$&$2.9$&$0.072917$&$0.03125$&$0.0022289$&$0.016893$&$0.0016$\\
\hdashline
$2$&$3$&$0.0665$&$0.026$&$0.00075946$&$0.016716$&$0.0016$\\
$2$&$4$&$0.0375$&$0.0083333$&$3.7777e-06$&$0.016235$&$000.11$\\
$3$&$4$&$0.041667$&$0.025$&$2.4814e-07$&$0.016712$&$0.0008$\\
$4$&$5$&$0.039583$&$0.03125$&$8.4841e-08$&$0.015931$&$0.0007$\\
\hline
 \end{tabular}
\caption{Values of the coupling masses which satisfy the WMAP bounds at the pivot scale $k_0=0.002\rm{Mpc}^{-1}$ and maximise the 
spectrum at $N=60$ $e-$folds. Also included is energy scale of the model
evaluated at $k=k_0$ in units of $\rm{\mpl}$.}
\end{table}

\begin{table}[h]
 \begin{tabular}{|c|c|c|c|c|c|c|}
\hline
$p$&$q$&$\eta_p$&$\eta_q$&$\calp_\zeta(k_{end})$&$n_s'(k_0)$&$V_0^{1/4}$\\
\hline
$2$&$2.2$&$0.16735$&$0.13878$&$0.20683$&$0.0049911$&$0.0008$\\
$2$&$2.3$&$0.12857$&$0.11633$&$0.15491$&$0.0067744$&$0.0009$\\
$2$&$2.3$&$0.096939$&$0.058163$&c$0.070402$&$0.010396$&$0.0012$\\
$2$&$2.7$&$0.085417$&$0.042$&$0.062686$&$0.01505$&$0.0016$\\
$2$&$2.9$&$0.071667$&$0.03125$&$0.0069787$&$0.016301$&$0.0016$\\
\hdashline
$2$&$3$&$0.066$&$0.026$&$0.0024599$&$0.016461$&$0.0016$\\
$2$&$4$&$0.036207$&$0.008$&$5.5689e-06$&$0.01517$&$0.0016$\\
$3$&$4$&$0.048276$&$0.034$&$3.1526e-07$&$0.016363$&$0.0017$\\
$4$&$5$&$0.026552$&$0.018$&$7.8237e-08$&$0.013356$&$0.0017$\\
\hline
 \end{tabular}
\caption{Values of the coupling masses which satisfy the WMAP bounds at the pivot scale $k_0=0.002\rm{Mpc}^{-1}$ and maximise the 
spectrum at $N=65$ $e-$folds. Also included is energy scale of the model
evaluated at $k=k_0$ in units of $\rm{\mpl}$.}
\end{table}

\section{The sensitivity curves}\label{app:sens_curves}

In this paper we plot the sensitivity curves of LIGO
\cite{ligo,adv_ligo},LISA \cite{lisa},  LCGT\cite{lcgt}, DECIGO
\cite{Seto:2001qf,decigo}.  For the space based detectors we use the
online sensitivity curve generator in \cite{sens_curves} inputting the
detector parameters  given in Table \ref{tab:sens}~\footnote{The $S_{\rm shot}$ values
for DECIGO and Ultimate DECIGO are the updated values of Ref.~\cite{Kudoh:2004he}
kindly provided by the authors in a private communication}. The sensitivity curve generator produces the strain
sensitivity defined as \cite{Cornish:2001bb},

\begin{table}
 \begin{tabular}{|c||c|c|c|c|c|c|}
  \hline
&$L[m]$&$S_{\rm shot}[\rm{mHz^{-1/2}}|]$&$S_{\rm accel}[\rm{ms^{-2}Hz^{-1/2}}]$&$P [W]$&$D[m]$&$\lambda[nm]$\\
\hline
LISA&$5\times10^9$&$2\times10^{-11}$&$3\times10^{-15}$&$1$&$0.3$&$1064$\\
BBO/DECIGO&$5\times10^7$&$1.1\times10^{-16}$&$7.9\times10^{-19}$&$10$&$1$&$532$\\
Ultimate DECIGO&$5\times10^7$&$1.7\times10^{-18}$&$3\times10^{-19}$&$10$&$1$&$532$\\
\hline
 \end{tabular}
\caption{Detector parameters used to input into the sensitivity curve generator in Ref.~\cite{sens_curves}.
Where $S_{\rm shot}$ is the root spectral position noise budget,
$S_{\rm accel}$ 
is the root spectral density acceleration noise, $P$ is the
power of the laser, $D$ is the telescope/mirror diameter and $\lambda$
is the wavelength. The new ESA only LISA mission has an arm length of $10^9~m$ \cite{AmaroSeoane:2012km}, and
the difference in sensitivity is shown in \fig{fig:lisa_elisa} in terms 
of the energy density of gravitational waves.}
\label{tab:sens}
\end{table}
\begin{figure}
 \centering\includegraphics[width=2.5in,totalheight=2.5in]{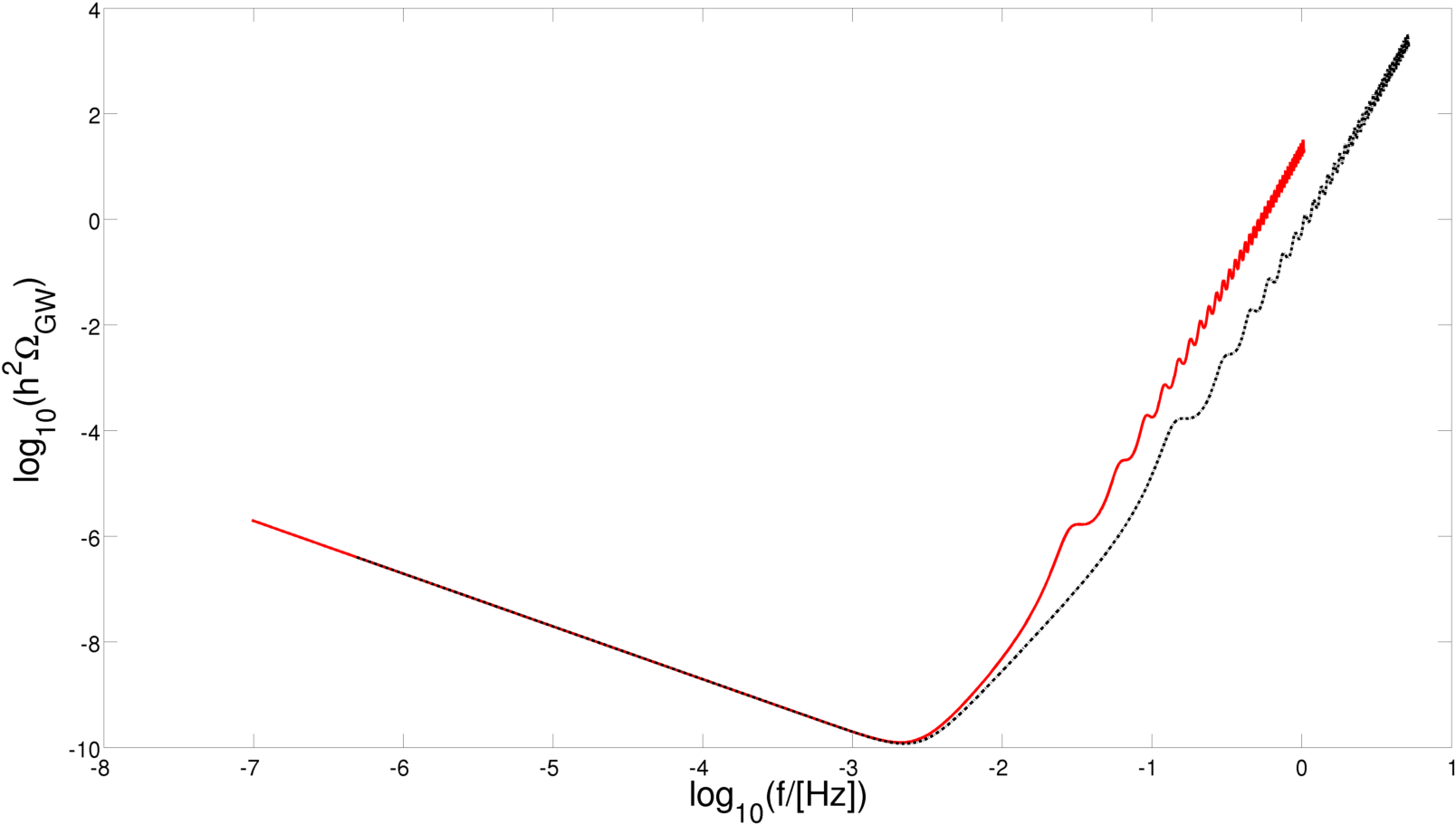}
\caption{Plot of the energy density of gravitational waves vs. the frequency in Hertz,
for the original NASA/ESA LISA mission (red) and the new ESA only eLISA mission
(black dash-dot). In this paper we have plotted the sensitivity range of the original LISA
mission, and as can be seen from this figure, the difference with the new eLISA mission
is negligible for our purposes.}
\label{fig:lisa_elisa}
\end{figure}

\be
h_{\rm eff}(f)=\sqrt{\frac{S_n(f)}{\mathcal{N}(f)}},
\ee
 where the units are in $Hz^{-1/2}$, and $S_n$ is known as the noise spectral density and $\mathcal{N}$ is the interferometer response function.
To convert between the strain efficiency and the energy density per logarithmic interval we use  
$\Omega_{gw}=4\pi^2f^3S_h(f)/(3H_0^2)$ \cite{Cornish:2001bb}, where $S_h$ is the signal spectral density
and we can then write this as \cite{Maggiore:2000gv} :

\be
h_0^2\Omega_{\rm GW}^{\rm min}(f)
\simeq0.0125\times{\rm{SNR}^2}\(\frac{f}{100\rm{Hz}}\)^3\(\frac{h_
{\rm eff}}{10^{-22}\rm{Hz}^{-1/2}}\)^2
\ee 
where $\rm{SNR}$ is the signal to noise ratio, which for interferometer 
type detectors is given as \cite{Maggiore:2000gv} ${\rm SNR}=\sqrt{2S_h(f)/S_n{f}}$.

Finally we also plot what is known as the \emph{cross-correlation} sensitivity of the BBO-DECIGO detector. In this
case two separate DECIGO detectors are launched and the correlation of their signals results in a significant
reduction in the overall sensitivity \cite{Kudoh:2005as}.
The strain efficiency of cross-correlated detectors is given as \cite{Cornish:2001bb}

\be
h_{\rm eff,cross}(f)=\frac{{\rm{SNR}^2}}{(2T\Delta{f})^{1/4}}\overline{\(\frac{|\mathcal{N}_{12}|^2}{S_{n1}S_{n2}}\)}^{(-1/4)}
\ee
where $T$ is the observation time in seconds, $\mathcal{N}_{12}$ is given as $<S_1S_2>=\int_0^\infty{}dfS_h(f)\mathcal{N}_{12}$,  $\Delta{}f$
is the frequency resolution, and the overbar is the average over a frequency interval. In this paper we use a rough estimate,
we assume that the
strain efficiency of each detector is the same and that the detectors are co-aligned and coincident (i.e. 
that their position vectors are the same). Then the strain efficiency reduces to

\be
h_{\rm eff,cross}\sim{}{\rm{SNR}^2}\frac{h_{\rm eff}}{(2T\Delta{}f)^{1/4}}
\ee
and in our analysis we take $\Delta{}f=f/10$ for 
the cross correlated DECIGO detector.

\bibliographystyle{JHEPmodplain}
\bibliography{SOGW}

\providecommand{\href}[2]{#2}\begingroup\raggedright\begin{thebibliography}{10}

\bibitem{Komatsu:2010fb}
{\bf WMAP} Collaboration, E.~Komatsu {\em et~al.}, {\it {Seven-Year Wilkinson
  Microwave Anisotropy Probe (WMAP) Observations: Cosmological
  Interpretation}},  {\sl Astrophys. J. Suppl.} {\bf 192} (2011) 18,
  [\href{http://arxiv.org/abs/1001.4538}{{\sf arXiv:1001.4538}}],
  [\href{http://dx.doi.org/10.1088/0067-0049/192/2/18}{{\sf
  doi:10.1088/0067-0049/192/2/18}}].

\bibitem{Abazajian:2008wr}
{\bf SDSS} Collaboration, K.~N. Abazajian {\em et~al.}, {\it {The Seventh Data
  Release of the Sloan Digital Sky Survey}},  {\sl Astrophys. J. Suppl.} {\bf
  182} (2009) 543--558, [\href{http://arxiv.org/abs/0812.0649}{{\sf
  arXiv:0812.0649}}],
  [\href{http://dx.doi.org/10.1088/0067-0049/182/2/543}{{\sf
  doi:10.1088/0067-0049/182/2/543}}].

\bibitem{Hicken:2009dk}
M.~Hicken, W.~Wood-Vasey, S.~Blondin, P.~Challis, S.~Jha, {\em et~al.}, {\it
  {Improved Dark Energy Constraints from ~100 New CfA Supernova Type Ia Light
  Curves}},  {\sl Astrophys.J.} {\bf 700} (2009) 1097--1140,
  [\href{http://arxiv.org/abs/0901.4804}{{\sf arXiv:0901.4804}}],
  [\href{http://dx.doi.org/10.1088/0004-637X/700/2/1097}{{\sf
  doi:10.1088/0004-637X/700/2/1097}}].

\bibitem{Meiksin:2007rz}
A.~A. Meiksin, {\it {The Physics of the Intergalactic Medium}},  {\sl
  Rev.Mod.Phys.} {\bf 81} (2009) 1405--1469,
  [\href{http://arxiv.org/abs/0711.3358}{{\sf arXiv:0711.3358}}],
  [\href{http://dx.doi.org/10.1103/RevModPhys.81.1405}{{\sf
  doi:10.1103/RevModPhys.81.1405}}].

\bibitem{Bird:2010mp}
S.~Bird, H.~V. Peiris, M.~Viel, and L.~Verde, {\it {Minimally Parametric Power
  Spectrum Reconstruction from the Lyman-alpha Forest}},  {\sl
  Mon.Not.Roy.Astron.Soc.} {\bf 413} (2011) 1717--1728,
  [\href{http://arxiv.org/abs/1010.1519}{{\sf arXiv:1010.1519}}],
  [\href{http://dx.doi.org/10.1111/j.1365-2966.2011.18245.x}{{\sf
  doi:10.1111/j.1365-2966.2011.18245.x}}]. * Temporary entry *.

\bibitem{Lewis:2006fu}
A.~Lewis and A.~Challinor, {\it {Weak gravitational lensing of the cmb}},  {\sl
  Phys.Rept.} {\bf 429} (2006) 1--65,
  [\href{http://arxiv.org/abs/astro-ph/0601594}{{\sf arXiv:astro-ph/0601594}}],
  [\href{http://dx.doi.org/10.1016/j.physrep.2006.03.002}{{\sf
  doi:10.1016/j.physrep.2006.03.002}}].

\bibitem{Komatsu:2002wc}
E.~Komatsu and U.~Seljak, {\it {The Sunyaev-Zel'dovich angular power spectrum
  as a probe of cosmological parameters}},  {\sl Mon.Not.Roy.Astron.Soc.} {\bf
  336} (2002) 1256, [\href{http://arxiv.org/abs/astro-ph/0205468}{{\sf
  arXiv:astro-ph/0205468}}],
  [\href{http://dx.doi.org/10.1046/j.1365-8711.2002.05889.x}{{\sf
  doi:10.1046/j.1365-8711.2002.05889.x}}].

\bibitem{Josan:2010vn}
A.~S. Josan and A.~M. Green, {\it {Gamma-rays from ultracompact minihalos:
  potential constraints on the primordial curvature perturbation}},  {\sl
  Phys.Rev.} {\bf D82} (2010) 083527,
  [\href{http://arxiv.org/abs/1006.4970}{{\sf arXiv:1006.4970}}],
  [\href{http://dx.doi.org/10.1103/PhysRevD.82.083527}{{\sf
  doi:10.1103/PhysRevD.82.083527}}].

\bibitem{Bringmann:2011ut}
T.~Bringmann, P.~Scott, and Y.~Akrami, {\it {Improved constraints on the
  primordial power spectrum at small scales from ultracompact minihalos}},
  \href{http://arxiv.org/abs/1110.2484}{{\sf arXiv:1110.2484}}.

\bibitem{Yang:2011eg}
Y.~Yang, L.~Feng, X.~Huang, X.~Chen, T.~Lu, {\em et~al.}, {\it {Constraints on
  ultracompact minihalos from extragalactic gamma-ray background}},  {\sl JCAP}
  {\bf 1112} (2011) 020, [\href{http://arxiv.org/abs/1112.6229}{{\sf
  arXiv:1112.6229}}].

\bibitem{Li:2012qh}
F.~Li, A.~L. Erickcek, and N.~M. Law, {\it {A new probe of the small-scale
  primordial power spectrum: astrometric microlensing by ultracompact
  minihalos}},  \href{http://arxiv.org/abs/1202.1284}{{\sf arXiv:1202.1284}}.

\bibitem{Scott:2012kx}
P.~Scott, T.~Bringmann, and Y.~Akrami, {\it {Constraints on small-scale
  cosmological perturbations from gamma-ray searches for dark matter}},
  \href{http://arxiv.org/abs/1205.1432}{{\sf arXiv:1205.1432}}.

\bibitem{Carr:2009jm}
B.~Carr, K.~Kohri, Y.~Sendouda, and J.~Yokoyama, {\it {New cosmological
  constraints on primordial black holes}},  {\sl Phys.Rev.} {\bf D81} (2010)
  104019, [\href{http://arxiv.org/abs/0912.5297}{{\sf arXiv:0912.5297}}],
  [\href{http://dx.doi.org/10.1103/PhysRevD.81.104019}{{\sf
  doi:10.1103/PhysRevD.81.104019}}].

\bibitem{Josan:2009qn}
A.~S. Josan, A.~M. Green, and K.~A. Malik, {\it {Generalised constraints on the
  curvature perturbation from primordial black holes}},  {\sl Phys.Rev.} {\bf
  D79} (2009) 103520, [\href{http://arxiv.org/abs/0903.3184}{{\sf
  arXiv:0903.3184}}], [\href{http://dx.doi.org/10.1103/PhysRevD.79.103520}{{\sf
  doi:10.1103/PhysRevD.79.103520}}].

\bibitem{Leach:2000ea}
S.~M. Leach, I.~J. Grivell, and A.~R. Liddle, {\it {Black hole constraints on
  the running mass inflation model}},  {\sl Phys. Rev.} {\bf D62} (2000)
  043516, [\href{http://arxiv.org/abs/astro-ph/0004296}{{\sf
  arXiv:astro-ph/0004296}}],
  [\href{http://dx.doi.org/10.1103/PhysRevD.62.043516}{{\sf
  doi:10.1103/PhysRevD.62.043516}}].

\bibitem{Kohri:2007gq}
K.~Kohri, C.-M. Lin, and D.~H. Lyth, {\it {More hilltop inflation models}},
  {\sl JCAP} {\bf 0712} (2007) 004, [\href{http://arxiv.org/abs/0707.3826}{{\sf
  arXiv:0707.3826}}],
  [\href{http://dx.doi.org/10.1088/1475-7516/2007/12/004}{{\sf
  doi:10.1088/1475-7516/2007/12/004}}].

\bibitem{Kohri:2007qn}
K.~Kohri, D.~H. Lyth, and A.~Melchiorri, {\it {Black hole formation and
  slow-roll inflation}},  {\sl JCAP} {\bf 0804} (2008) 038,
  [\href{http://arxiv.org/abs/0711.5006}{{\sf arXiv:0711.5006}}],
  [\href{http://dx.doi.org/10.1088/1475-7516/2008/04/038}{{\sf
  doi:10.1088/1475-7516/2008/04/038}}].

\bibitem{Alabidi:2009bk}
L.~Alabidi and K.~Kohri, {\it {Generating Primordial Black Holes Via
  Hilltop-Type Inflation Models}},  {\sl Phys. Rev.} {\bf D80} (2009) 063511,
  [\href{http://arxiv.org/abs/0906.1398}{{\sf arXiv:0906.1398}}],
  [\href{http://dx.doi.org/10.1103/PhysRevD.80.063511}{{\sf
  doi:10.1103/PhysRevD.80.063511}}].

\bibitem{Josan:2010cj}
A.~S. Josan and A.~M. Green, {\it {Constraints from primordial black hole
  formation at the end of inflation}},  {\sl Phys.Rev.} {\bf D82} (2010)
  047303, [\href{http://arxiv.org/abs/1004.5347}{{\sf arXiv:1004.5347}}],
  [\href{http://dx.doi.org/10.1103/PhysRevD.82.047303}{{\sf
  doi:10.1103/PhysRevD.82.047303}}].

\bibitem{Matarrese:1993zf}
S.~Matarrese, O.~Pantano, and D.~Saez, {\it {General relativistic dynamics of
  irrotational dust: Cosmological implications}},  {\sl Phys.Rev.Lett.} {\bf
  72} (1994) 320--323, [\href{http://arxiv.org/abs/astro-ph/9310036}{{\sf
  arXiv:astro-ph/9310036}}],
  [\href{http://dx.doi.org/10.1103/PhysRevLett.72.320}{{\sf
  doi:10.1103/PhysRevLett.72.320}}].

\bibitem{Matarrese:1997ay}
S.~Matarrese, S.~Mollerach, and M.~Bruni, {\it {Second order perturbations of
  the Einstein-de Sitter universe}},  {\sl Phys.Rev.} {\bf D58} (1998) 043504,
  [\href{http://arxiv.org/abs/astro-ph/9707278}{{\sf arXiv:astro-ph/9707278}}],
  [\href{http://dx.doi.org/10.1103/PhysRevD.58.043504}{{\sf
  doi:10.1103/PhysRevD.58.043504}}].

\bibitem{Noh:2004bc}
H.~Noh and J.-c. Hwang, {\it {Second-order perturbations of the Friedmann world
  model}},  {\sl Phys.Rev.} {\bf D69} (2004) 104011,
  [\href{http://dx.doi.org/10.1103/PhysRevD.69.104011}{{\sf
  doi:10.1103/PhysRevD.69.104011}}].

\bibitem{Carbone:2004iv}
C.~Carbone and S.~Matarrese, {\it {A Unified treatment of cosmological
  perturbations from super-horizon to small scales}},  {\sl Phys.Rev.} {\bf
  D71} (2005) 043508, [\href{http://arxiv.org/abs/astro-ph/0407611}{{\sf
  arXiv:astro-ph/0407611}}],
  [\href{http://dx.doi.org/10.1103/PhysRevD.71.043508}{{\sf
  doi:10.1103/PhysRevD.71.043508}}].

\bibitem{Nakamura:2004rm}
K.~Nakamura, {\it {Second-order gauge invariant cosmological perturbation
  theory: Einstein equations in terms of gauge invariant variables}},  {\sl
  Prog.Theor.Phys.} {\bf 117} (2007) 17--74,
  [\href{http://arxiv.org/abs/gr-qc/0605108}{{\sf arXiv:gr-qc/0605108}}],
  [\href{http://dx.doi.org/10.1143/PTP.117.17}{{\sf doi:10.1143/PTP.117.17}}].
  Complete version of gr-qc/0605107.

\bibitem{Ananda:2006af}
K.~N. Ananda, C.~Clarkson, and D.~Wands, {\it {The Cosmological gravitational
  wave background from primordial density perturbations}},  {\sl Phys.Rev.}
  {\bf D75} (2007) 123518, [\href{http://arxiv.org/abs/gr-qc/0612013}{{\sf
  arXiv:gr-qc/0612013}}],
  [\href{http://dx.doi.org/10.1103/PhysRevD.75.123518}{{\sf
  doi:10.1103/PhysRevD.75.123518}}].

\bibitem{Baumann:2007zm}
D.~Baumann, P.~J. Steinhardt, K.~Takahashi, and K.~Ichiki, {\it {Gravitational
  Wave Spectrum Induced by Primordial Scalar Perturbations}},  {\sl Phys.Rev.}
  {\bf D76} (2007) 084019, [\href{http://arxiv.org/abs/hep-th/0703290}{{\sf
  arXiv:hep-th/0703290}}],
  [\href{http://dx.doi.org/10.1103/PhysRevD.76.084019}{{\sf
  doi:10.1103/PhysRevD.76.084019}}].

\bibitem{Cook:2011hg}
J.~L. Cook and L.~Sorbo, {\it {Particle production during inflation and
  gravitational waves detectable by ground-based interferometers}},  {\sl
  Phys.Rev.} {\bf D85} (2012) 023534,
  [\href{http://arxiv.org/abs/1109.0022}{{\sf arXiv:1109.0022}}],
  [\href{http://dx.doi.org/10.1103/PhysRevD.85.023534}{{\sf
  doi:10.1103/PhysRevD.85.023534}}]. 20 pages, 3 figures, version on Phys. Rev.
  D.

\bibitem{Assadullahi:2009jc}
H.~Assadullahi and D.~Wands, {\it {Constraints on primordial density
  perturbations from induced gravitational waves}},  {\sl Phys.Rev.} {\bf D81}
  (2010) 023527, [\href{http://arxiv.org/abs/0907.4073}{{\sf
  arXiv:0907.4073}}], [\href{http://dx.doi.org/10.1103/PhysRevD.81.023527}{{\sf
  doi:10.1103/PhysRevD.81.023527}}].

\bibitem{lisa}
\url{http://lisa.nasa.gov/}.

\bibitem{Seto:2001qf}
N.~Seto, S.~Kawamura, and T.~Nakamura, {\it {Possibility of direct measurement
  of the acceleration of the universe using 0.1-Hz band laser interferometer
  gravitational wave antenna in space}},  {\sl Phys.Rev.Lett.} {\bf 87} (2001)
  221103, [\href{http://arxiv.org/abs/astro-ph/0108011}{{\sf
  arXiv:astro-ph/0108011}}],
  [\href{http://dx.doi.org/10.1103/PhysRevLett.87.221103}{{\sf
  doi:10.1103/PhysRevLett.87.221103}}].

\bibitem{decigo}
\url{http://tamago.mtk.nao.ac.jp/decigo/index_E.html}.

\bibitem{Mollerach:2003nq}
S.~Mollerach, D.~Harari, and S.~Matarrese, {\it {CMB polarization from
  secondary vector and tensor modes}},  {\sl Phys.Rev.} {\bf D69} (2004)
  063002, [\href{http://arxiv.org/abs/astro-ph/0310711}{{\sf
  arXiv:astro-ph/0310711}}],
  [\href{http://dx.doi.org/10.1103/PhysRevD.69.063002}{{\sf
  doi:10.1103/PhysRevD.69.063002}}].

\bibitem{Saito:2009jt}
R.~Saito and J.~Yokoyama, {\it {Gravitational-Wave Constraints on the Abundance
  of Primordial Black Holes}},  {\sl Prog.Theor.Phys.} {\bf 123} (2010)
  867--886, [\href{http://arxiv.org/abs/0912.5317}{{\sf arXiv:0912.5317}}],
  [\href{http://dx.doi.org/10.1143/PTP.123.867}{{\sf
  doi:10.1143/PTP.123.867}}]. * Brief entry *.

\bibitem{Bugaev:2009kq}
E.~Bugaev and P.~Klimai, {\it {Bound on induced gravitational wave background
  from primordial black holes}},  {\sl JETP Lett.} {\bf 91} (2010) 1--5,
  [\href{http://arxiv.org/abs/0911.0611}{{\sf arXiv:0911.0611}}],
  [\href{http://dx.doi.org/10.1134/S0021364010010017}{{\sf
  doi:10.1134/S0021364010010017}}]. * Brief entry *.

\bibitem{Bugaev:2010bb}
E.~Bugaev and P.~Klimai, {\it {Constraints on the induced gravitational wave
  background from primordial black holes}},  {\sl Phys.Rev.} {\bf D83} (2011)
  083521, [\href{http://arxiv.org/abs/1012.4697}{{\sf arXiv:1012.4697}}],
  [\href{http://dx.doi.org/10.1103/PhysRevD.83.083521}{{\sf
  doi:10.1103/PhysRevD.83.083521}}].

\bibitem{Bugaev:2009zh}
E.~Bugaev and P.~Klimai, {\it {Induced gravitational wave background and
  primordial black holes}},  {\sl Phys.Rev.} {\bf D81} (2010) 023517,
  [\href{http://arxiv.org/abs/0908.0664}{{\sf arXiv:0908.0664}}],
  [\href{http://dx.doi.org/10.1103/PhysRevD.81.023517}{{\sf
  doi:10.1103/PhysRevD.81.023517}}].

\bibitem{Bullock:1996at}
J.~S. Bullock and J.~R. Primack, {\it {NonGaussian fluctuations and primordial
  black holes from inflation}},  {\sl Phys.Rev.} {\bf D55} (1997) 7423--7439,
  [\href{http://arxiv.org/abs/astro-ph/9611106}{{\sf arXiv:astro-ph/9611106}}],
  [\href{http://dx.doi.org/10.1103/PhysRevD.55.7423}{{\sf
  doi:10.1103/PhysRevD.55.7423}}].

\bibitem{Ivanov:1997ia}
P.~Ivanov, {\it {Nonlinear metric perturbations and production of primordial
  black holes}},  {\sl Phys.Rev.} {\bf D57} (1998) 7145--7154,
  [\href{http://arxiv.org/abs/astro-ph/9708224}{{\sf arXiv:astro-ph/9708224}}],
  [\href{http://dx.doi.org/10.1103/PhysRevD.57.7145}{{\sf
  doi:10.1103/PhysRevD.57.7145}}].

\bibitem{PinaAvelino:2005rm}
P.~Pina~Avelino, {\it {Primordial black hole constraints on non-gaussian
  inflation models}},  {\sl Phys.Rev.} {\bf D72} (2005) 124004,
  [\href{http://arxiv.org/abs/astro-ph/0510052}{{\sf arXiv:astro-ph/0510052}}],
  [\href{http://dx.doi.org/10.1103/PhysRevD.72.124004}{{\sf
  doi:10.1103/PhysRevD.72.124004}}].

\bibitem{Chongchitnan:2006wx}
S.~Chongchitnan and G.~Efstathiou, {\it {Accuracy of slow-roll formulae for
  inflationary perturbations: implications for primordial black hole
  formation}},  {\sl JCAP} {\bf 0701} (2007) 011,
  [\href{http://arxiv.org/abs/astro-ph/0611818}{{\sf arXiv:astro-ph/0611818}}],
  [\href{http://dx.doi.org/10.1088/1475-7516/2007/01/011}{{\sf
  doi:10.1088/1475-7516/2007/01/011}}].

\bibitem{Hidalgo:2007vk}
J.~Hidalgo, {\it {The effect of non-Gaussian curvature perturbations on the
  formation of primordial black holes}},
  \href{http://arxiv.org/abs/0708.3875}{{\sf arXiv:0708.3875}}.

\bibitem{Bugaev:2011wy}
E.~Bugaev and P.~Klimai, {\it {Formation of primordial black holes from
  non-Gaussian perturbations produced in a waterfall transition}},
  \href{http://arxiv.org/abs/1112.5601}{{\sf arXiv:1112.5601}}. v2: 11 pages, 4
  figures. Several comments and references added. Version accepted by Phys.
  Rev. D.

\bibitem{Chluba}
J.~Chluba, A.~L. Erickcek, and I.~Ben-Dayan, {\it {Probing the inflaton:
  Small-scale power spectrum constraints from measurements of the CMB energy
  spectrum}},  \href{http://arxiv.org/abs/1203.2681}{{\sf arXiv:1203.2681}}.

\bibitem{Lyth:2009zz}
D.~H. Lyth and A.~R. Liddle, {\it {The primordial density perturbation:
  Cosmology, inflation and the origin of structure}}, .

\bibitem{Drees:2011yz}
M.~Drees and E.~Erfani, {\it {Running Spectral Index and Formation of
  Primordial Black Hole in Single Field Inflation Models}},
  \href{http://arxiv.org/abs/1110.6052}{{\sf arXiv:1110.6052}}.

\bibitem{Allahverdi:2006cx}
R.~Allahverdi, A.~Kusenko, and A.~Mazumdar, {\it {A-term inflation and the
  smallness of neutrino masses}},  {\sl JCAP} {\bf 0707} (2007) 018,
  [\href{http://arxiv.org/abs/hep-ph/0608138}{{\sf arXiv:hep-ph/0608138}}],
  [\href{http://dx.doi.org/10.1088/1475-7516/2007/07/018}{{\sf
  doi:10.1088/1475-7516/2007/07/018}}].

\bibitem{Lin:2008ys}
C.-M. Lin and K.~Cheung, {\it {Super Hilltop Inflation}},  {\sl JCAP} {\bf
  0903} (2009) 012, [\href{http://arxiv.org/abs/0812.2731}{{\sf
  arXiv:0812.2731}}],
  [\href{http://dx.doi.org/10.1088/1475-7516/2009/03/012}{{\sf
  doi:10.1088/1475-7516/2009/03/012}}].

\bibitem{Lin:2009yt}
C.-M. Lin and K.~Cheung, {\it {Reducing the Spectral Index in Supernatural
  Inflation}},  {\sl Phys.Rev.} {\bf D79} (2009) 083509,
  [\href{http://arxiv.org/abs/0901.3280}{{\sf arXiv:0901.3280}}],
  [\href{http://dx.doi.org/10.1103/PhysRevD.79.083509}{{\sf
  doi:10.1103/PhysRevD.79.083509}}].

\bibitem{Kohri:2010sj}
K.~Kohri and C.-M. Lin, {\it {Hilltop Supernatural Inflation and Gravitino
  Problem}},  {\sl JCAP} {\bf 1011} (2010) 010,
  [\href{http://arxiv.org/abs/1008.3200}{{\sf arXiv:1008.3200}}],
  [\href{http://dx.doi.org/10.1088/1475-7516/2010/11/010}{{\sf
  doi:10.1088/1475-7516/2010/11/010}}].

\bibitem{Hotchkiss:2011gz}
S.~Hotchkiss, A.~Mazumdar, and S.~Nadathur, {\it {Observable gravitational
  waves from inflation with small field excursions}},
  \href{http://arxiv.org/abs/1110.5389}{{\sf arXiv:1110.5389}}.

\bibitem{Stewart:1996ey}
E.~D. Stewart, {\it {Flattening the inflaton's potential with quantum
  corrections}},  {\sl Phys.Lett.} {\bf B391} (1997) 34--38,
  [\href{http://arxiv.org/abs/hep-ph/9606241}{{\sf arXiv:hep-ph/9606241}}],
  [\href{http://dx.doi.org/10.1016/S0370-2693(96)01458-X}{{\sf
  doi:10.1016/S0370-2693(96)01458-X}}].

\bibitem{Covi:1998jp}
L.~Covi, D.~H. Lyth, and L.~Roszkowski, {\it {Observational constraints on an
  inflation model with a running mass}},  {\sl Phys. Rev.} {\bf D60} (1999)
  023509, [\href{http://arxiv.org/abs/hep-ph/9809310}{{\sf
  arXiv:hep-ph/9809310}}],
  [\href{http://dx.doi.org/10.1103/PhysRevD.60.023509}{{\sf
  doi:10.1103/PhysRevD.60.023509}}].

\bibitem{Covi:1998mb}
L.~Covi and D.~H. Lyth, {\it {Running-mass models of inflation, and their
  observational constraints}},  {\sl Phys. Rev.} {\bf D59} (1999) 063515,
  [\href{http://arxiv.org/abs/hep-ph/9809562}{{\sf arXiv:hep-ph/9809562}}],
  [\href{http://dx.doi.org/10.1103/PhysRevD.59.063515}{{\sf
  doi:10.1103/PhysRevD.59.063515}}].

\bibitem{Lyth:2000qp}
D.~H. Lyth and L.~Covi, {\it {Observational constraints on the spectral index
  of the cosmological curvature perturbation}},  {\sl Phys. Rev.} {\bf D62}
  (2000) 103504, [\href{http://arxiv.org/abs/astro-ph/0002397}{{\sf
  arXiv:astro-ph/0002397}}],
  [\href{http://dx.doi.org/10.1103/PhysRevD.62.103504}{{\sf
  doi:10.1103/PhysRevD.62.103504}}].

\bibitem{Covi:2000qx}
L.~Covi and D.~H. Lyth, {\it {Global fits for the spectral index of the
  cosmological curvature perturbation}},  {\sl Mon. Not. Roy. Astron. Soc.}
  {\bf 326} (2001) 885, [\href{http://arxiv.org/abs/astro-ph/0008165}{{\sf
  arXiv:astro-ph/0008165}}],
  [\href{http://dx.doi.org/10.1046/j.1365-8711.2001.04466.x}{{\sf
  doi:10.1046/j.1365-8711.2001.04466.x}}].

\bibitem{Covi:2002th}
L.~Covi, D.~H. Lyth, and A.~Melchiorri, {\it {New constraints on the
  running-mass inflation model}},  {\sl Phys. Rev.} {\bf D67} (2003) 043507,
  [\href{http://arxiv.org/abs/hep-ph/0210395}{{\sf arXiv:hep-ph/0210395}}],
  [\href{http://dx.doi.org/10.1103/PhysRevD.67.043507}{{\sf
  doi:10.1103/PhysRevD.67.043507}}].

\bibitem{Covi:2004tp}
L.~Covi, D.~H. Lyth, A.~Melchiorri, and C.~J. Odman, {\it {The running-mass
  inflation model and WMAP}},  {\sl Phys. Rev.} {\bf D70} (2004) 123521,
  [\href{http://arxiv.org/abs/astro-ph/0408129}{{\sf arXiv:astro-ph/0408129}}],
  [\href{http://dx.doi.org/10.1103/PhysRevD.70.123521}{{\sf
  doi:10.1103/PhysRevD.70.123521}}].

\bibitem{Frampton:2010sw}
P.~H. Frampton, M.~Kawasaki, F.~Takahashi, and T.~T. Yanagida, {\it {Primordial
  Black Holes as All Dark Matter}},  {\sl JCAP} {\bf 1004} (2010) 023,
  [\href{http://arxiv.org/abs/1001.2308}{{\sf arXiv:1001.2308}}],
  [\href{http://dx.doi.org/10.1088/1475-7516/2010/04/023}{{\sf
  doi:10.1088/1475-7516/2010/04/023}}].

\bibitem{Kawasaki:2012kn}
M.~Kawasaki, A.~Kusenko, and T.~T. Yanagida, {\it {Primordial seeds of
  supermassive black holes}},  \href{http://arxiv.org/abs/1202.3848}{{\sf
  arXiv:1202.3848}}.

\bibitem{num_rec}
{\em Numerical Recipes, Third Edition}.
\newblock Cambridge University Press, 2007.

\bibitem{ligo}
\url{http://www.ligo.caltech.edu/~jzweizig/distribution/LSC_Data/}.

\bibitem{lcgt}
\url{http://gwcenter.icrr.u-tokyo.ac.jp/en/}.

\bibitem{AmaroSeoane:2012km}
P.~Amaro-Seoane, S.~Aoudia, S.~Babak, P.~Binetruy, E.~Berti, {\em et~al.}, {\it
  {eLISA: Astrophysics and cosmology in the millihertz regime}},
  \href{http://arxiv.org/abs/1201.3621}{{\sf arXiv:1201.3621}}.

\bibitem{Jenet}
F.~A. Jenet, G.~Hobbs, W.~van Straten, R.~Manchester, M.~Bailes, {\em et~al.},
  {\it {Upper bounds on the low-frequency stochastic gravitational wave
  background from pulsar timing observations: Current limits and future
  prospects}},  {\sl Astrophys.J.} {\bf 653} (2006) 1571--1576,
  [\href{http://arxiv.org/abs/astro-ph/0609013}{{\sf arXiv:astro-ph/0609013}}],
  [\href{http://dx.doi.org/10.1086/508702}{{\sf doi:10.1086/508702}}].

\bibitem{Hobbs}
G.~Hobbs, D.~Miller, R.~Manchester, J.~Dempsey, J.~Chapman, {\em et~al.}, {\it
  {The Parkes Observatory Pulsar Data Archive}},
  \href{http://arxiv.org/abs/1105.5746}{{\sf arXiv:1105.5746}}.

\bibitem{Yardley}
D.~Yardley, G.~Hobbs, F.~Jenet, J.~Verbiest, Z.~Wen, {\em et~al.}, {\it {The
  Sensitivity of the Parkes Pulsar Timing Array to Individual Sources of
  Gravitational Waves}},  \href{http://arxiv.org/abs/1005.1667}{{\sf
  arXiv:1005.1667}}.

\bibitem{PPTA}
\url{http://www.atnf.csiro.au/research/pulsar/ppta/index.php?n=Main.PPTA}.

\bibitem{Boyle:2005ug}
L.~A. Boyle, P.~J. Steinhardt, and N.~Turok, {\it {Inflationary predictions
  reconsidered}},  {\sl Phys.Rev.Lett.} {\bf 96} (2006) 111301,
  [\href{http://arxiv.org/abs/astro-ph/0507455}{{\sf arXiv:astro-ph/0507455}}],
  [\href{http://dx.doi.org/10.1103/PhysRevLett.96.111301}{{\sf
  doi:10.1103/PhysRevLett.96.111301}}].

\bibitem{Chongchitnan:2006pe}
S.~Chongchitnan and G.~Efstathiou, {\it {Prospects for direct detection of
  primordial gravitational waves}},  {\sl Phys.Rev.} {\bf D73} (2006) 083511,
  [\href{http://arxiv.org/abs/astro-ph/0602594}{{\sf arXiv:astro-ph/0602594}}],
  [\href{http://dx.doi.org/10.1103/PhysRevD.73.083511}{{\sf
  doi:10.1103/PhysRevD.73.083511}}].

\bibitem{Kuroyanagi:2010mm}
S.~Kuroyanagi, T.~Chiba, and N.~Sugiyama, {\it {Prospects for Direct Detection
  of Inflationary Gravitational Waves by Next Generation Interferometric
  Detectors}},  {\sl Phys.Rev.} {\bf D83} (2011) 043514,
  [\href{http://arxiv.org/abs/1010.5246}{{\sf arXiv:1010.5246}}],
  [\href{http://dx.doi.org/10.1103/PhysRevD.83.043514}{{\sf
  doi:10.1103/PhysRevD.83.043514}}].

\bibitem{Joyce:1996cp}
M.~Joyce, {\it {Electroweak Baryogenesis and the Expansion Rate of the
  Universe}},  {\sl Phys.Rev.} {\bf D55} (1997) 1875--1878,
  [\href{http://arxiv.org/abs/hep-ph/9606223}{{\sf arXiv:hep-ph/9606223}}],
  [\href{http://dx.doi.org/10.1103/PhysRevD.55.1875}{{\sf
  doi:10.1103/PhysRevD.55.1875}}].

\bibitem{Seto:2003kc}
N.~Seto and J.~Yokoyama, {\it {Probing the equation of state of the early
  universe with a space laser interferometer}},  {\sl J.Phys.Soc.Jap.} {\bf 72}
  (2003) 3082--3086, [\href{http://arxiv.org/abs/gr-qc/0305096}{{\sf
  arXiv:gr-qc/0305096}}], [\href{http://dx.doi.org/10.1143/JPSJ.72.3082}{{\sf
  doi:10.1143/JPSJ.72.3082}}].

\bibitem{Boyle:2007zx}
L.~A. Boyle and A.~Buonanno, {\it {Relating gravitational wave constraints from
  primordial nucleosynthesis, pulsar timing, laser interferometers, and the
  CMB: Implications for the early Universe}},  {\sl Phys.Rev.} {\bf D78} (2008)
  043531, [\href{http://arxiv.org/abs/0708.2279}{{\sf arXiv:0708.2279}}],
  [\href{http://dx.doi.org/10.1103/PhysRevD.78.043531}{{\sf
  doi:10.1103/PhysRevD.78.043531}}].

\bibitem{Hidalgo:2011fj}
J.~Hidalgo, L.~A. Urena-Lopez, and A.~R. Liddle, {\it {Unification models with
  reheating via Primordial Black Holes}},  {\sl Phys.Rev.} {\bf D85} (2012)
  044055, [\href{http://arxiv.org/abs/1107.5669}{{\sf arXiv:1107.5669}}],
  [\href{http://dx.doi.org/10.1103/PhysRevD.85.044055}{{\sf
  doi:10.1103/PhysRevD.85.044055}}]. Updated to match version accepted by PRD.

\bibitem{Nakayama:2008ip}
K.~Nakayama, S.~Saito, Y.~Suwa, and J.~Yokoyama, {\it {Space laser
  interferometers can determine the thermal history of the early Universe}},
  {\sl Phys.Rev.} {\bf D77} (2008) 124001,
  [\href{http://arxiv.org/abs/0802.2452}{{\sf arXiv:0802.2452}}],
  [\href{http://dx.doi.org/10.1103/PhysRevD.77.124001}{{\sf
  doi:10.1103/PhysRevD.77.124001}}].

\bibitem{Nakayama:2008wy}
K.~Nakayama, S.~Saito, Y.~Suwa, and J.~Yokoyama, {\it {Probing reheating
  temperature of the universe with gravitational wave background}},  {\sl JCAP}
  {\bf 0806} (2008) 020, [\href{http://arxiv.org/abs/0804.1827}{{\sf
  arXiv:0804.1827}}],
  [\href{http://dx.doi.org/10.1088/1475-7516/2008/06/020}{{\sf
  doi:10.1088/1475-7516/2008/06/020}}].

\bibitem{Acquaviva:2002ud}
V.~Acquaviva, N.~Bartolo, S.~Matarrese, and A.~Riotto, {\it {Second order
  cosmological perturbations from inflation}},  {\sl Nucl.Phys.} {\bf B667}
  (2003) 119--148, [\href{http://arxiv.org/abs/astro-ph/0209156}{{\sf
  arXiv:astro-ph/0209156}}],
  [\href{http://dx.doi.org/10.1016/S0550-3213(03)00550-9}{{\sf
  doi:10.1016/S0550-3213(03)00550-9}}].

\bibitem{Baskaran:2006qs}
D.~Baskaran, L.~Grishchuk, and A.~Polnarev, {\it {Imprints of Relic
  Gravitational Waves in Cosmic Microwave Background Radiation}},  {\sl
  Phys.Rev.} {\bf D74} (2006) 083008,
  [\href{http://arxiv.org/abs/gr-qc/0605100}{{\sf arXiv:gr-qc/0605100}}],
  [\href{http://dx.doi.org/10.1103/PhysRevD.74.083008}{{\sf
  doi:10.1103/PhysRevD.74.083008}}].

\bibitem{Dodelson:2003ft}
S.~Dodelson, {\it {Modern cosmology}}, .
  \url{http://home.fnal.gov/~dodelson/book.html}.

\bibitem{abr}
M.~{Abramowitz} and I.~A. {Stegun}, {\em Handbook of Mathematical Functions
  with Formulas, Graphs, and Mathematical Tables}.
\newblock Dover, New York, ninth dover printing, tenth gpo printing~ed., 1964.

\bibitem{Maggiore:2000gv}
M.~Maggiore, {\it {Stochastic backgrounds of gravitational waves}},
  \href{http://arxiv.org/abs/gr-qc/0008027}{{\sf arXiv:gr-qc/0008027}}.

\bibitem{adv_ligo}
\url{https://dcc.ligo.org/cgi-bin/DocDB/ShowDocument?docid=2974}.

\bibitem{sens_curves}
\url{http://www.srl.caltech.edu/~shane/sensitivity/}.

\bibitem{Kudoh:2004he}
H.~Kudoh and A.~Taruya, {\it {Probing anisotropies of gravitational-wave
  backgrounds with a space-based interferometer: Geometric properties of
  antenna patterns and their angular power}},  {\sl Phys.Rev.} {\bf D71} (2005)
  024025, [\href{http://arxiv.org/abs/gr-qc/0411017}{{\sf
  arXiv:gr-qc/0411017}}],
  [\href{http://dx.doi.org/10.1103/PhysRevD.71.024025}{{\sf
  doi:10.1103/PhysRevD.71.024025}}].

\bibitem{Cornish:2001bb}
N.~J. Cornish, {\it {Detecting a stochastic gravitational wave background with
  the Laser Interferometer Space Antenna}},  {\sl Phys.Rev.} {\bf D65} (2002)
  022004, [\href{http://arxiv.org/abs/gr-qc/0106058}{{\sf
  arXiv:gr-qc/0106058}}],
  [\href{http://dx.doi.org/10.1103/PhysRevD.65.022004}{{\sf
  doi:10.1103/PhysRevD.65.022004}}].

\bibitem{Kudoh:2005as}
H.~Kudoh, A.~Taruya, T.~Hiramatsu, and Y.~Himemoto, {\it {Detecting a
  gravitational-wave background with next-generation space interferometers}},
  {\sl Phys.Rev.} {\bf D73} (2006) 064006,
  [\href{http://arxiv.org/abs/gr-qc/0511145}{{\sf arXiv:gr-qc/0511145}}],
  [\href{http://dx.doi.org/10.1103/PhysRevD.73.064006}{{\sf
  doi:10.1103/PhysRevD.73.064006}}].

\end{thebibliography}\endgroup

\end{document}